\documentclass[11pt]{article}
\usepackage[english]{babel}
\pdfoutput=1
\usepackage{amsmath}
\usepackage{amssymb}
\usepackage[section]{placeins}
\usepackage{graphicx}
\usepackage[left=2cm,right=2cm,top=2cm,bottom=2cm]{geometry}

\newcommand{\ave}[1]{\left\langle #1\right\rangle}

\newcommand{\rood}[1]{}

\begin{document}

\title{Geometrically-Consistent Model Reduction of Polymer Chains in Solution. Application to Dissipative Particle Dynamics: Model Description} 

\author{Nicolas Moreno$^{1,2}$, Suzana P. Nunes$^1$ and Victor M. Calo$^{2,3}$}
\date{}

\maketitle 
\begin{center}
$^1$ Environmental Science and Engineering, Water Desalination and Reuse Center,  KAUST, $^2$ Center for Numerical Porous Media NUMPOR, KAUST, $^3$ Applied Mathematics \& Computational Science,  Earth Science \& Engineering, KAUST
\end{center}

\begin{abstract}
We introduce a framework for model reduction of chain models for dissipative particle dynamics (DPD) simulations, where the characteristic size of the chain, pressure, density,  and temperature are preserved. The proposed methodology reduces the number of degrees of freedom required to represent a particular system with complex molecules (e.g., linear polymers). Based on geometrical considerations we map fine-grained models to a reference state through a consistent scaling of the system, where short length and  fast time scales are disregarded while the properties governing the phase equilibria are preserved. Following this coarse-graining process we consistently represent high molecular weight DPD chains (i.e., $\geq 200$ beads per chain) with a significant reduction in the number of particles required (i.e., $\geq 20$ times the original system). 
\end{abstract}

\section{Introduction}

Dissipative particle dynamics (DPD) is a stochastic mesoscale particle model introduced by Hoogerbrugge and Koelman.\cite{Search1992} DPD combines features from
molecular dynamics (MD) and lattice-gas automata (LGA) to simulate the isothermal Navier-Stokes (NS) equations. The resulting method is faster than MD and avoids
the lattice artifacts of LGA. Espa\~nol and Warren\cite{Search1995}
reformulated the DPD method, describing it within the statistical mechanics framework.
\\
\\
One of the most important application of DPD is the study of polymers,\cite{Espa2005, Cao2005}
amphiphiles,\cite{Li2009, Amsterdam1999, Shillcock2012, Yamamoto2002} and their mixtures. Other applications include hydrodynamic and
excluded volume interactions,\cite{Jiang2007} collapse transitions going from good
to poor solvents,\cite{Kong1997, Zhao2013} rheological properties,\cite{Espa2005} self assembly of diblock copolymers in solution,\cite{Horsch2004, He2010, Spaeth2011a, Marques2013} and microphase separation.\cite{Groot1998, Chen2013}
\\
\\
DPD models complex material behavior through the interactions of soft
particles (a.k.a., beads). Beads are typically described as a single point with a soft repulsive interaction
potential which has a cut-off radius $r_c$.
In DPD liquids are modeled by single interacting beads, while
polymers (or any complex structure)  can be simply constructed joining many DPD
particles through bonding potentials such as harmonic springs. Polymer solutions with different concentrations are modeled changing
the ratio between the number of polymer and solvent beads. Furthermore, the solvent quality can be varied by fine tuning the  solvent-solvent
and solvent-polymer interaction parameters used to set up DPD simulations. 
\\
\\
One of the most important limitations in the modeling of high molecular weight molecules is that the large polymer chains need to be discretely represented in the model as well as their interaction with the solvent.\rood{PENDING:MAYBE ADD THE DEFAULT CG 3WATER PER BEAD AS EXAMPLE AS ICCS PAPER.} This large number of beads leads to a high computational demand which limits the time and length scales attainable.\cite{Tromov2003, Brennan2009} In this scenario, the modeling of polymers in practical applications ranging from infinitely dilute solutions to self-assembly is still cumbersome. Different authors \cite{Backer2005, Fuchslin2009} have proposed methodologies to reduce the number of particles needed to describe a DPD system with fluids and polymers chains.\cite{Spaeth2011} However the applicability of these methodologies for polymeric systems is restricted to short chains.\cite{Spaeth2011, Moreno2014a}
\\
\\
Motivated by the current limitations in the modeling of arbitrary long chain models, herein we describe a methodology to reduce the total number of degrees of freedom necessary to accurately forecast the behavior of complex molecules. We use the term \textit{particle} or \textit{bead} to refer to the degrees of freedom in our simulation, while \textit{segment} denotes the particles that constitute a chain.  Due to the meso-scale nature of DPD this segments can be associated for example with monomers, Kuhn segments or blobs, when physical systems are translated to DPD.  
\\
\\
This paper is organized as follows. First, we introduce the conventional DPD governing equations as well as the conformational characterization we use for DPD chains. Then, we present the proposed model reduction framework. In the remaining sections we present the validation of the coarse-graining introduced and draw conclusions.

\section{Dissipative particle dynamics}

In DPD the kinematic evolution and the balance of linear momentum of the particles are given by
\\
\begin{align} 
\frac{d\textbf{r}_i}{dt} &=\textbf{v}_i,
\\
m_i\frac{d\textbf{v}_i}{dt} &=\textbf{f}_i = \sum_{j \neq i}(\textbf{F}_{ij}^C+\textbf{F}_{ij}^D+\textbf{F}_{ij}^R),
\end{align}
where $\textbf{r}_i$, $\textbf{v}_i$ are the position and velocity of a
particle $i$, respectively, $m_i$ is its mass, and $\textbf{f}_i$
is the net force acting over the particle. The force acting on each
particle, has three different contributions, $\textbf{F}_{ij}^C$ is a conservative force, that models pressure effects between particles and spring interactions in chain models. $\textbf{F}_{ij}^D$, models dissipative (viscous) interactions
in a fluid (a friction force that reduces the velocity
differences between particles). $\textbf{F}_{ij}^R$ is a random force
(stochastic) that models random collisions between particles, and from
the MD point of view, models the degrees of freedom eliminated by the
coarse-graining process. This stochastic force approximates the Brownian motion of
polymers and colloids. From the statistical mechanics point of view,
$\textbf{F}_{ij}^D$ and $\textbf{F}_{ij}^R$ are tightly related in order to satisfy the fluctuation-dissipation theorem, which takes the  form of the Fokker-Plank equation. \cite{Search1995}
\\
\\
The conservative force typically can be written as $\textbf{F}_{ij}^C = \textbf{F}_{ij}^{B} + \textbf{F}_{ij}^{S}$, where $\textbf{F}_{ij}^{B}$ and $\textbf{F}_{ij}^{S}$ account for bead-bead and bead-spring (when particles are connected) interactions, respectively.\cite{Posel2014} In terms of their energy potentials $u_{ij}$, the bead-bead and bead-spring contributions can be expressed as
\begin{align}
\textbf{F}_{ij}^B &= -\frac{\text{d}u_{ij}^B}{\text{dr}_{ij}}\frac{\textbf{r}_{ij}}{|r_{ij}|}, 
\\
\textbf{F}_{ij}^S &=   -\delta_{ij}\frac{\text{d}u_{ij}^S}{\text{dr}_{ij}}\frac{\textbf{r}_{ij}}{|r_{ij}|},
\end{align}
where $\textbf{r}_{ij} = \textbf{r}_i - \textbf{r}_j$ and $r_{ij} = |\textbf{r}_{ij}|$.  $\delta_{ij} = 1$ if particles $i$ and $j$ are connected, and $\delta_{ij} = 0$ otherwise. In the literature the most used bead-spring energy potentials are harmonic and finite-extensible-non-linear elastic springs,\cite{Symeonidis2005, Posel2014} however other alternatives are possible.\cite{Symeonidis2005} Regarding the bead-bead contribution, soft-repulsive potentials are typically chosen as the simplest option,\cite{Groot1997} nevertheless more rigorous potentials can be used.\cite{Pagonabarraga2001} The model reduction framework we propose can be applied to any form of the conservative force adopted.   
\\
\\
The remaining forces are defined as
\begin{equation}
\textbf{F}_{ij}^D=-\gamma \omega ^D(r_{ij})\left(\frac{\textbf{r}_{ij}}{|r_{ij}|}\cdot
\textbf{v}_{ij}\right)\frac{\textbf{r}_{ij}}{|r_{ij}|},
\end{equation}
\begin{equation}
\textbf{F}_{ij}^R=\sigma \omega ^R(r_{ij}) \zeta \Delta t^{-1/2}
\frac{\textbf{r}_{ij}}{|r_{ij}|},
\end{equation} 
where $\gamma$ is a friction coefficient that determines the overall
magnitude of the dissipative term, and $\sigma$ is the noise amplitude that
scales the stochastic contribution. $\omega ^D$ and $\omega ^R$  are weighting functions that set the range of interaction between particles. $\zeta$ is a random number with zero mean and unit variance; the dependence of $\textbf{F}_{ij}^R$ with the time step size, appears as an important restriction in the time integration procedure. The different forces satisfy Newton's third law, and conserve linear and angular momenta. 
\\
\\
According to Espanol and Warren,\cite{Search1995} the system satisfies a
Gaussian distribution only if
\\
\begin{equation}
\omega^D(r_{ij}) = [\omega^R(r_{ij})]^2,
\end{equation}
similarly, from the fluctuation-dissipation theorem, the noise
amplitude and the dissipative coefficient are related by
\\
\begin{equation}
\sigma^2 = 2\gamma k_BT,
\end{equation}
where $k_B$ is the Boltzmann constant and $T$ is the equilibrium temperature.
Due to its simplicity, the following definition for the weighting function $\omega^R(r_{ij})$ (and therefore $\omega^D(r_{ij})$) is commonly used in the literature
\\
\begin{equation}
\omega^D(r_{ij}) =[\omega^R(r_{ij})]^2=\left\{\begin{array}{rl}
(1-r_{ij}/r_c)^2; & (r_{ij}<r_c), \\ 
0; &(r_{ij}\geq r_c),
\end{array}\right. 
\end{equation}
where $\omega^R(r_{ij})$ is assumed to vary linearly away from the particle.
\subsection{Conformational Characterization of linear polymer chains}


The equilibrium distribution of beads	 along the DPD chains and therefore the chain size is in general governed by the enthalpic and entropic interactions
between segments. The entropic contribution can be associated with the
configuration of the polymer, such as linear, star, branched, etc. While the enthalpic contributions are in general governed by
the polymer-polymer and/or polymer-solvent  interactions in solution.  
\\
\\
To motivate this discussion we present our coarse grain methodology in the context of linear polymer configurations. Nevertheless, the methodology we present can be further applied to other chain configurations. 
\\
\\
Here, a linear polymer is defined as a sequence of $N + 1$ particles connected, with an equilibrium length between them of $r_{o} = br_c$, where $b$ is a proportionality constant. We use the traditional
polymer chain distinction between \textit{ideal} (or \textit{theta}) and \textit{real}
chains conformations.\cite{Rubinstein2003}
A chain is in its ideal configuration when there are no energetic interactions between segments, or the balance between interactions cancel each other (i.e., theta condition). Thus, any particle only interacts with those particles it is directly connected to. In this ideal state the segments nearly behave as in a random-walk distribution, or self-avoiding walk in theta condition. A polymer chain with conformations different from ideal is assumed to exhibit a real configuration.   
\\
\\
%
In order to characterize the size of the DPD chain model we use three non-zero measurements,\cite{Rubinstein2003} the mean-square radius $\ave{{R}^2}$, the radius of gyration $R_g$ and the contour length $l_c$.  The ensemble average over configuration is denoted by $\left\langle \cdot \right\rangle$. The mean-square radius $\ave{{R}^2}$ is given by 

\begin{equation}
\ave{R^2} = \ave{\textbf{R}_N \cdot \textbf{R}_N} = \sum_{i=1}^N \sum_{j=1}^N \ave{\textbf{s}_i \cdot \textbf{s}_j}.
\end{equation} 
where $\textbf{R}_N$ is the end-to-end vector, and $\textbf{s}_i$ is the bond vector pointing from the $(i-1)$th to the $i$th
segment in the chain. We can express $\ave{\textbf{s}_i \cdot \textbf{s}_j} = \ave{r_{ij,o}^2~\cos \theta_{ij}}$, where $\theta_{ij}$ is the angle between
$\textbf{s}_i$ and $\textbf{s}_j$ and $r_{ij,o}$ is the distance between particles. If the  distance between connecting particles is assumed almost uniform, that is, $\ave{r_{ij,o}} \approx r_o$, the mean-squared radius can be rewritten 

\begin{equation}
\langle R^2 \rangle =  r_o^2 \left(\sum_{i=1}^{N} \sum_{j=1}^{N} \langle \cos \theta_{ij} \rangle \right).
\label{eq:meansquaredistance}
\end{equation}
The magnitude of $\cos \theta_{ij}$ measures the orientation  similarity
between the vectors $\textbf{s}_i$ and $\textbf{s}_j$. This
measurement is commonly known as cosine similarity. That is, $\ave{\cos \theta_{ij}} $ provides relevant information about the
correlation between segments; for ideal chain models $\langle R^2 \rangle =  r_o^2N$ because there is
no correlation between segments $\ave{\cos \theta_{ij}}=0$ if $i\neq
j$. 
%
However, for real chains 

\begin{equation}
\ave{\cos \theta_{ij}} \neq 0,  \quad \quad \text{for} |i-j|<\varepsilon.
\label{eq:cosnoideal}
\end{equation}
Here $\varepsilon$ indicates the segment separation where the correlations
between segments vanishes. To express the mean-square
radius in a more generic form for both ideal and real chain conformations, we
introduce the so called  Flory's characteristic ratio,\cite{Rubinstein2003} $C_N$ 

\begin{equation}
C_N = \frac{1}{N} \sum_{i=1}^N C_i,
\end{equation}
where
\begin{equation}
C_i = \sum_{j =1}^N \left\langle \cos \theta_{ij} \right\rangle.
\end{equation}
Now, using the definition of Flory's characteristic ratio we obtain

\begin{equation}
\ave{R^2} = r_o^2NC_N.
\label{eq:rsquare}
\end{equation}
In polymer physics the numerical value of
$C_N$ depends on the local stiffness of the polymer chain. For some
polymers the correlation between monomers separated by many bonds
disappears, and the Flory's correlation function saturates to a value
$C_{\infty}$.\cite{Rubinstein2003}
\\
\\
Using $C_N$ we characterize the
conformation and correlation between segments in our DPD chain
models. From \eqref{eq:rsquare}  we conclude that $C_N=1$ for ideal chains, and $C_N=N$ for rod-shaped chains
(completely extended).  Therefore, if we write 

\begin{align}
C_N \approx N^{\beta},
\label{eq:floryratio}
\end{align}
and $\nu = (1+\beta)/2$, the mean-square radius of the chain can be written as a power law $\ave{R^2} = r_o^2 N^{2\nu}$, therefore,
\begin{equation}
|R| = \sqrt{\ave{R^2}} = r_o N^{\nu},
\label{eq:aveR}
\end{equation}
where $\nu$ provides information about the chain conformation (segment correlation), and depends on the  affinity between the DPD chain and the
\textit{surrounding particles}. Here the surrounding particles
account for the chain concentration effects on the system. Thus,
in diluted systems the DPD chains are mostly surrounded by solvent
particles, but as the chain concentration increases the chains start
interacting with other chains.
\\
\\
Different authors \cite{SchlijperAGandHoogerbruggePJandManke1995a,
Kong1997, Spenley2000, Symeonidis2005, Yang2013, Ilnytskyi2007, Posel2014} have shown that in DPD spatial and temporal correlations appear and the power laws underlying polymer
physics \cite{Rubinstein2003} can be captured with DPD polymer chains.  It has been verified experimentally \cite{Rubinstein2003} that $1/3 \leq \nu \leq 1$ depending on the chain size and the solvent affinity. In general, all the polymers have values of $\nu$ that fall in this range, irrespective of the concentration regime, that is, for diluted, semidiluted, concentrated or bulk. Nevertheless, the chain size and solvent affinity at which each $\nu$ value is achieved is specific to each polymer analysed. 
\\
\\
The values of $\nu$ in the range $1/3$ to $1$ can be interpreted geometrically and this sheds light on the correlation between beads in the chain. A brief discussion of this geometrical interpretation is given in the following. Equation \eqref{eq:aveR} shows that the radius of
a sphere that contains the polymer chain grows proportionally to some power $\nu$ of the
number of segments $N$.  For $\nu=1/3$, equation \eqref{eq:aveR} implies that
$R^3 \approx r_o^3N = \sum^N r_o^3$. Therefore the volume of the sphere is
approximately the summation of the volumes of the $N$
segments, since the segments are tightly packed.  A single polymer chain in poor solvent exhibiting $\nu = 1/3$ is expected to be completely collapsed. 
In contrast, at the largest value $\nu=1$, the radius of the sphere
containing the polymer scales as $R \approx r_oN = \sum^N r_o$, therefore the only
possible segment configuration is a completely extended chain. In summary at $\nu \approx 1/3$ the polymer chain packing behaves like a sphere while at $\nu \approx 1$ the polymer chains behaves like a rod.
\\
\\
These geometrical considerations explain the limits of
$\nu$, passing from fully collapsed to fully extended chain
arrangements. The segments are
assumed to be incompressible; otherwise $R^3 < r_o^3N$ when the chain is
collapsed or  $R > r_o N$ when it is extended.
\\
\\
Another useful measure we use to characterize a polymer chain model is the radius of gyration $R_g$, defined as

\begin{equation}
R_g^2 = \frac{1}{N}\sum_{i=1}^N \left\langle (\textbf{r}_i-\textbf{r}_{cm})^2 \right\rangle = \frac{1}{N^2}\sum_{i=1}^N\sum_{j=i}^N \left\langle (\textbf{r}_i-\textbf{r}_{j})^2 \right\rangle.
\end{equation}
where $\textbf{r}_i$ is the position vector of the $i$th particle. The radius of gyration corresponds to the second moment around the center of mass
for the segments position $\textbf{r}_{cm}$ in a polymer chain. In
general $R_g \propto R$, particularly if  the chain exhibits the same conformation at all
scales, it is possible to integrate over the polymer contour,\cite{Rubinstein2003} leading to a general expression for any $\nu$, which is

\begin{equation}
\left\langle R_g^2 \right\rangle = r_o^2 \frac{N^{2\nu}}{(2\nu+1)(2\nu+2)} = \frac{\ave{R^2}}{(2\nu+1)(2\nu+2)}.
\label{eq:rggeneral}
\end{equation}
From \eqref{eq:rggeneral}, we conclude that the radius of gyration for ideal chains
is defined as $\left\langle R_g^2 \right\rangle = Nr_o^2/6$, while for
rod-shape structures, $R_g^2 = {N^2r_o^2}/{12}$.
\\
\\
The last parameter we use to characterize the size of a DPD chain is the \textit{contour length} $l_c$, which we defined as

\begin{equation}
\ave{l_c} = \sum_{i=1}^N \ave{|\textbf{s}_i|}  \approx r_o N.
\label{eq:contour}
\end{equation}
\section{Coarse-Graining process}
\label{sec:coarsegraining}

\begin{figure}[h!]
\centering
\includegraphics[width=0.75\linewidth]{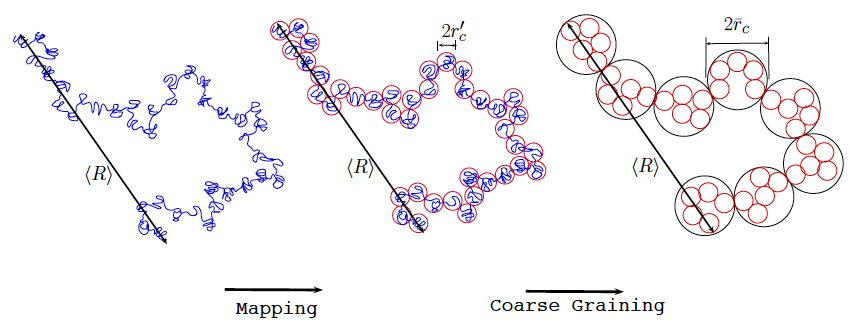}

\caption{Schematic representation of the process of mapping and coarse graining of a polymer chain.}
\label{fig:mappingAndCG}
\end{figure}

In this section, we describe the framework we propose to reduce the number of degrees of freedom needed to accurately forecast the behavior of complex molecules. In this context we introduce a distinction between:
\begin{itemize}
\item [\textit{i}.] the process of fitting physical properties with DPD parameters, that we call \textit{mapping} and,
\item[\textit{ii}.] the process of reducing the number of degrees of freedom of a given DPD system, that we call \textit{coarse graining} or \textit{model reduction}.
\end{itemize} 
The mapping process in itself requires a coarse graining procedure, where physical atoms are grouped in DPD-particle representations, Thus, model reduction is simply a particular type of coarse graining during the mapping process (Figure \ref{fig:mappingAndCG}). However for the sake of clarity we prefer to introduce mapping and coarse graining to focus our efforts on proposing a general methodology for model reduction, that can be combined with any mapping from the literature, or used in the development of more sophisticated mapping procedures.  
\\
\\

In this paper the \textit{coarse graining} process reduces the number of particles that describes a given system by grouping these into coarse sets (Figure \ref{fig:mappingAndCG}).  Prior to coarse graining, the particles in the
system are labelled as the \textit{fine} particle representation, and
after coarse graining, we identify the system components as \textit{coarse}
particles. In order to systematically present the proposed coarse graining
methodology we distinguish between the values of a property $A$ when evaluated on the fine grained system ($A'$) and its coarse grained counterpart ($\bar{A}$).
\\
\\
The ratio  between the number of particles before and after the model reduction is the \textit{level of coarse graining} $\phi$, which is a measure of the reduction in the number of degrees of freedom representing the system. We define the level of coarse graining based on the change in the number of particles used to represent the polymer chain, such that

\begin{equation}
\phi = \frac{{N'}}{\bar{N}}.
\label{eq:lofcg}
\end{equation}
We assume that this level of coarse graining is applied to the whole system (i.e., solvent and polymer), the total number of particles $N_T$ in a coarse system is given by

\begin{equation}
\bar{N}_T  = \phi {N'}_T.
\end{equation}
In order to preserve particular features of the fully resolved system (e.g., pressure $p$, mass density $\rho$) once the coarse graining is applied, the DPD parameters of  the coarse system need to be properly adjusted. We call this procedure parameter \textit{scaling}. The value of a given parameter $\bar{A}$ is computed by scaling $A'$ as

\begin{equation}
\bar{A} = \psi(\phi)A',
\end{equation}
where $\psi(\phi)$ is a scaling function that depends on the level of coarse graining.
\\
\\
One of the first attempts to formalize model reduction in DPD using
scaling arguments was presented by Backer et al.\cite{Backer2005} In their work the authors described a methodology to scale DPD parameters in
flow problems using particles with two different cutoff radii
(multiresolution).
Later, Fuchslin et al.,\cite{Fuchslin2009} introduced a scaling scheme to deal with fluid-like systems containing individual particles,
that consistently coarse grain DPD, and
restated it as a scale-free mesoscopic method, thus it can be applied to any
lenght scale.  
\\
\\
Following the approach of Backer et al.,\cite{Backer2005}, Spaeth et al.,\cite{Spaeth2011} extended the methodology for arbitrary coarse graining level
and introduced the same idea for coarse
graining of polymer chains. As Backer,\cite{Backer2005} they scaled the parameters in order to
preserve the mass density $\rho$, the pressure of the system $p$,
the number of interactions per particle, and the viscosity
$\mu$. The model reduction proposed by Spaeth
\cite{Spaeth2011} preserves satisfactorily
the target features of the original fine system only up to certain
maximum chain length ($40$ beads/chain) and coarse graining ($\phi=5$). This suggests the
existence of additional features related with the chain length that are
not being accounted properly when the DPD parameters are scaled using this framework.  
\\
\\
In the original model for combined scales proposed by Backer,
\cite{Backer2005} and adopted by Spaeth\cite{Spaeth2011} the
cutoff radius is scaled in order to preserve the mass density and the
number of interactions per particle of the original system. That is, they assumed that a  coarse particle represents a set of $n$ fine particles homogeneously distributed within the coarse volume.
This assumption is valid if the coarse grained particles do not
exhibit non-local correlations between them (i.e., fluids represented by
individual particles), however for polymer chains
correlations between connected particles exists, and depending on the
polymer shape, long range correlations may also be relevant. Hence, the particle correlations in the chain representations seems to be responsible of the maximum limit in coarse graining and chain length reported by Spaeth et al.\cite{Spaeth2011}
\\
\\
We propose a methodology for coarse-graining DPD-chains where the correlation between particles is
explicitly included and preserved by the model reduction framework.
The coarse-graining approach proposed generalizes the
seminal ideas of Fuchslin,\cite{Fuchslin2009} including relevant concepts from polymer
physics such as power laws.\cite{Rubinstein2003} Our coarse graining by construction preserves the most relevant features of the fine grained system, in particular, the characteristic size $R$ (a.k.a., end-to-end distance $R_f$) of the DPD chain. Nevertheless, due to its practical relevance we also verify the preservation of the radius of gyration in the coarse grained models.

\subsection{Mass, $m$, and cutoff radius, $r_c$, scaling}

\begin{figure}[h!] 
\begin{center}
\includegraphics[width=0.75\linewidth]{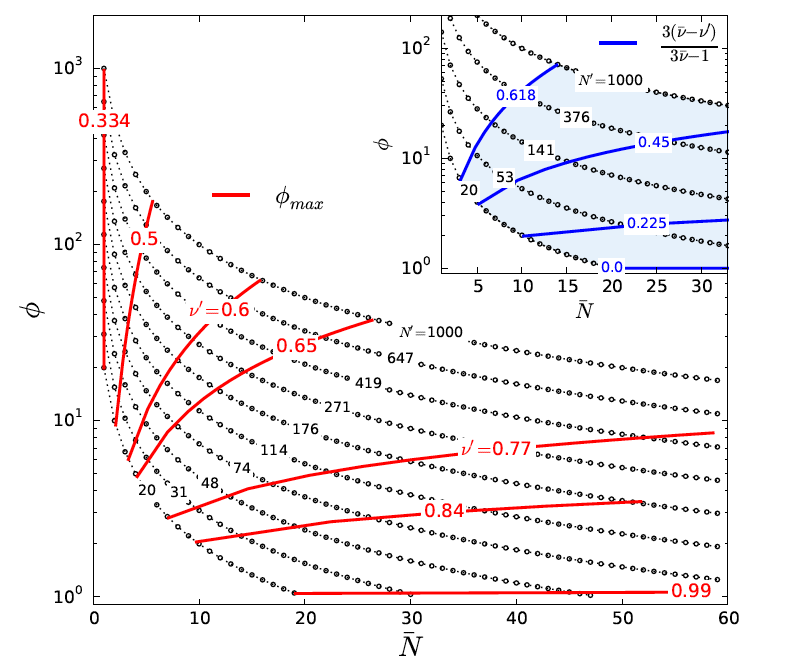}
\end{center}
\caption{Given a fine-scale chain with $N'$ beads, the number of particles $\bar{N}$ of its coarse-grained counterpart is restricted by the fine and coarse conformations $\nu'$ and $\bar{\nu}$, respectively. The continuous lines are the maximum level of coarse graining that can be achieved for fine scale chains with different conformations $\nu'$.  The dotted curves indicates the level of coarse graining for fine-scale chains ranging from 20 to 1000 beads. Since the number of particles per chain must be an integer, each dot corresponds to discrete values of $\phi$. The inset Figure represents the variation of the coarse-graining level with $\bar{\nu}$ for fine-scale chains with $\nu'=0.6$.}
\label{fig:phiandNbarvariation}
\end{figure}

We define the mass $m$ of a coarse-grained particle as

\begin{equation}
\bar{m} = \phi m'.
\label{eq:scalemass}
\end{equation}
The unit length in a system constituted only by fine particles is given
by the cutoff radius $r_{c}'$. We scale the cutoff radius to preserve the proper chain dimension $R = {r'}_oN'^{\nu'}$. Since the average distance between connected particles is proportional to the cutoff radius, $r_o = {b} r_c$, in order to preserve the radial particle distribution the proportionality constant ${b}$ must be the same for any coarse-graining level. Hence, if we attempt to preserve the characteristic polymer size $R$ after coarse graining, we require that

\begin{align}
R = N'^{\nu'} {b} r_{c}' = \bar{N}^{\bar{\nu}} {b} \bar{r}_{c},
\label{eq:scaleRadius}
\end{align}
therefore the cutoff radius of the system with coarse particles is given by
\begin{equation}
\bar{r}_{c} = r_{c}'~N'^{\nu' - \bar{\nu}}\phi^{\bar{\nu}},
\label{eq:rcscaling}
\end{equation}
where the scaling proposed in \cite{Spaeth2011} is recovered when $\bar{\nu} = \nu' = 1/3$. Thus, the maximum chain length and coarse graining limitation that the method of \cite{Spaeth2011} suffers is due to the implicit assumption that the coarse and fine models have identical spatial correlations. Moreover, $\nu = 1/3$ assumes a
complete collapse of the chain, which is only valid in the poor-solvent
limit.
\\
\\
Now, if the mass density is to be preserved in the coarse representation too, then

\begin{equation}
\rho = \frac{m'}{v'} = \frac{\bar{m}}{\bar{v}},
\label{eq:rhomass}
\end{equation}
where $m$ and $v$ are the mass and volume per particle, respectively. Based on the unit length scaling \eqref{eq:rcscaling}, the volume of a coarse particle is given by

\begin{align}
\bar{v} &=  \frac{4\pi}{3}(r_{c}')^3~(N'^{(\nu' - \bar{\nu})}\phi^{\bar{\nu}})^3 =  v'~(N'^{(\nu' - \bar{\nu})}\phi^{\bar{\nu}})^3.
\label{eq:massandvol}
\end{align}
Again, equation \eqref{eq:massandvol} for $\nu' = \bar{\nu} =
1/3$, collapses to the scaling scheme of \cite{Spaeth2011} ($\bar{v} = \phi v'$). Substituting  \eqref{eq:scalemass} $\bar{m}= m' \phi$ and \eqref{eq:massandvol}  in \eqref{eq:rhomass}, we get,

\begin{equation}
\phi = N'^{\frac{3(\bar{\nu} - \nu')}{3\bar{\nu} - 1}}.
\label{eq:sigmaandnu}
\end{equation}
In \eqref{eq:sigmaandnu} we have one free parameter, that is, either $\phi$ or
$\bar{\nu}$. Thus once one of them has been chosen the other is fixed by \eqref{eq:sigmaandnu} 
if the mass density is going to be preserved by the coarse graining process.
Here we choose the coarse graining parameter $\phi$ as the free variable, leading to

\begin{equation}
\bar{\nu} = \frac{1}{3}~\frac{3\nu' \text{ln}N' - \text{ln}\phi}{\text{ln}N' - \text{ln}\phi}.
\label{eq:nu1}
\end{equation}
In principle, from \eqref{eq:sigmaandnu} $1 \leq \phi \leq N'$ (equation \eqref{eq:lofcg}), however
due to the physical restriction that $1/3 \leq \bar{\nu} \leq 1$ (assuming
incompressibility of the segments), the level of coarse graining must exhibit a $\phi_{max}$ when $\bar{\nu} = 1$. From Figure \eqref{eq:sigmaandnu} $\phi_{max}$ can be easily computed. The maximum coarse graining for a polymer chain with $N'$ segments and $\nu'$ is given by

\begin{equation}
\phi_{max} = N'^{(3/2)(1-\nu')}.
\label{eq:sigmax}
\end{equation}
In Figure \ref{fig:phiandNbarvariation} we present the level of coarse graining required to represent fine-grained chains as shorter coarse-grained model. Additionally in this Figure we include the maximum level of coarse graining \eqref{eq:sigmax} attainable for different fine scale conformations $\nu'$. The existence of a maximum coarse-graining level arises from the fact
that we want to preserve both mass density and polymer size. From \ref{fig:phiandNbarvariation} it is evident that for $\nu'  \approx 1$
(rod-shape polymer),  $\phi_{max}=1$, therefore the chain cannot be
coarsened while preserving the mass and length scale (equations \eqref{eq:scaleRadius} and \eqref{eq:rhomass}).
\\
\\
In the inset of Figure \ref{fig:phiandNbarvariation} we highlight the variation of the level of coarse graining for different values of the exponent of equation \eqref{eq:sigmaandnu}. This Figure is useful to identify graphically the magnitude of $\bar{\nu}$. In this case the blue region denotes the permissible levels of coarse graining, for chains with fine-scale conformation $\nu'=0.6$. This region is delimited by $\phi_{max}$, when $\bar{\nu}=1$, and $\phi = 1$, when $\bar{\nu} = 0.6$. We remark the fact that if the fine and coarse representation have the same conformation, the chain cannot be reduced any longer. 

\subsection{Time $\tau$ and energy $\epsilon$ scalings}

In DPD the time scale is given by
\begin{equation}
\tau^2 =  r_c^2\frac{m}{\epsilon},
\label{eq:tau}
\end{equation}
where the traditional selection of $r_c=1$, $m=1$ and $\epsilon = k_BT =
1$ leads to a DPD time scale of size one. In this case the time scale is
defined once the energy units are chosen. $\epsilon = k_BT$ is
adequate in the study of equilibrium states, however the time scale
can also be determined by direct comparison of experimental and
simulated transport coefficients.\cite{DZWINEL2000, Fuchslin2009, Fu2013}  
\\
\\
Since our goal is to perform model reduction on DPD while preserving relevant simulations parameters close to the original fine-scale simulation, the scaling of the mass \eqref{eq:scalemass}  and length \eqref{eq:scaleRadius} units stated implies an appropriate
scaling of the units of time and energy. In our coarse
graining we adopt the scaling of time $\psi_{time}(\phi)$ and
energy $\psi_{energy}(\phi)$ proposed in \cite{Fuchslin2009} based on dimensional analysis of
\eqref{eq:tau}. Therefore, we have for the coarse system that

\begin{align}
\bar{\tau}^2 \bar{\epsilon} &= \bar{r_{c}}^2\bar{m},
\end{align}
which can be expanded to yield
\begin{align}
(\tau' \psi_{time}(\phi))^2 \epsilon' \psi_{energy}(\phi) &=  r_{c}'^2~N'^{2(\nu' - \bar{\nu})}\phi^{2\bar{\nu}} m' \phi, 
\label{eq:tauScaling}
\end{align}  
where the choice of $\psi_{time}(\phi)$ and $\psi_{energy}(\phi)$ depends on
the scaling parameters acting on the cutoff radius and mass. Based on
\eqref{eq:tauScaling}, a simple alternative is to scale the unit of energy
as the particle mass is scaled and the unit of time as the cutoff
radius. Therefore 

\begin{align}
\psi_{time}(\phi) &= N'^{(\nu' - \bar{\nu})}\phi^{\bar{\nu}},
\\
\psi_{energy}(\phi) &= \phi.
\label{eq:energyScaling}
\end{align}
Even though the selection of the energy and time scalings are
arbitrary, \eqref{eq:energyScaling} is useful in order to
preserve the physical scalability of DPD. \cite{Fuchslin2009}

\subsection{Parameter scaling for conservative interactions}
  
The scaling of the conservative interactions follows the methodology introduced by Fuchslin et. al.,\cite{Fuchslin2009} based on internal energy considerations. In \cite{Fuchslin2009}  the authors adopt a conservative contribution that only depends of soft-repulsive bead-bead interactions, herein we first generalize their approach and then particularize it to the bead-bead and bead-spring potentials frequently used in the literature. 
\\
\\
According to \cite{Fuchslin2009} the conservative interaction parameters scale in order to preserve the change of the internal energy ($\Delta U$) when the system is isotropically compressed from a box size $L$ to $(1-\zeta)L$, where $\zeta \ll 1$ is the relative compression parameter. The change of the internal energy in the system can be written as

\begin{align}
\Delta U &= U_{\zeta} - U_0,
\\
			&= \sum_{i=1}^{N_T} \sum_{j>i}^{N_T} u_{ij}(r_{ij} - \Delta r_{ij}(\zeta)) - u_{ij}(r_{ij}),
\end{align}
where $U_0$ is the internal energy of the uncompressed system, and $\Delta r_{ij}(\zeta)$ is the change in the distance between particles. 
\\
\\
The typical energy potentials used in the literature can be compactly written as $u_{ij} = \mathcal{A}h(r_{ij})$, where $\mathcal{A}$ is a constant that dictates the magnitude of the internal energy, while $h(r_{ij})$ sets the extent and order of the particles interactions. The scaling of the conservative contributions is applied to $\mathcal{A}$ such that the change in the internal energy $\Delta U$ can be preserved after model reduction. Thus,  if $\Delta U' =  \Delta \bar{U}$, we require that
\begin{align}
\sum_{i=1}^{N'_T} \sum_{j>i}^{N'_T} \mathcal{A}' \left( h({r'}_{ij} - \Delta {r'}_{ij}(\zeta)) - h({r'}_{ij}) \right) = \nonumber
\\
 \sum_{i=1}^{\bar{N}_T} \sum_{j>i}^{\bar{N}_T} \bar{\mathcal{A}} \left( h(\bar{r}_{ij} - \Delta \bar{r}_{ij}(\zeta)) - h(\bar{r}_{ij}) \right).
\label{eq:scalePotential}
\end{align}
In equation \eqref{eq:scalePotential} $\bar{\mathcal{A}}$ must compensate any change occurred in the right-hand side, such that the equality holds.  The changes in the right-hand-side expression are associated with a reduction in the number of terms in the summation (i.e., $N_T$) and the increment in the length scale (i.e., $r_c$). If we use the asymptotic behavior of the two contributions (i.e., when only one of the contribution dominates), and consider that both effects are uncoupled, the scaling of the conservative contributions can be expressed as

\begin{equation}
\bar{\mathcal{A}} = \psi_c(\phi) \mathcal{A}' = \psi_{c,\hat{\rho}}(\phi)\psi_{c,r_c}(\phi)\mathcal{A}',
\label{eq:scaleConservative}
\end{equation} 
where $\psi_{c,\hat{\rho}}(\phi)$ and $\psi_{c,r_c}(\phi)$ are the scaling functions due to the changes in the particle number density ($\hat{\rho} = N_T/V$) and length scale, respectively. Due to the short-range nature of DPD  and that the number of interactions per particle is conserved after coarse graining, the number of terms in the inner-most summation does not change. This is true for bead-bead interactions when the normalized radial distribution function does not depend on the level of coarse graining, $g({r'}_{ij}/{r'}_c) = g(\bar{r}_{ij}/\bar{r}_c)$, as we show in section \ref{sec:validation}. In bead-spring potentials, the number of interactions per particle only changes if the number of bonds per bead is modified with the coarse graining, which is not the case for the linear polymers discussed in this paper.
\\
\\
If we initially consider only the change in the number of particles $N_T$, and $h({r'}_{ij}) = h(\bar{r}_{ij})$ equation \eqref{eq:scalePotential} becomes $\mathcal{A}'N'_{T} = \bar{\mathcal{A}} \bar{N}_T$, leading to

\begin{equation}
\bar{\mathcal{A}} = \frac{N'_T}{\bar{N}_T} \mathcal{A}' = \phi \mathcal{A}' = \psi_{c,\hat{\rho}} \mathcal{A}'.
\label{eq:scalebyN}
\end{equation}
The effect of the change in the length scale can be identified when $N'_T = \bar{N}_T$, which from equation \eqref{eq:scalePotential} yields an scaling
\begin{equation}
\bar{\mathcal{A}} = \frac{h({r'}_{ij} - \Delta {r'}_{ij}(\zeta)) - h({r'}_{ij})}{h(\bar{r}_{ij} - \Delta \bar{r}_{ij}(\zeta)) - h(\bar{r}_{ij})} \mathcal{A}' = \psi_{c,r_c} \mathcal{A}'.
\label{eq:scalebyrc}
\end{equation}
\\
\\
Once the functional form of the bead-bead and bead-spring  potentials is chosen, the proper scaling of the conservative contributions can be determined from equations \eqref{eq:scalebyN} and \eqref{eq:scalebyrc}. To illustrate this point we choose the classical bead-bead potential used by \cite{Backer2005}  and \cite{Fuchslin2009}, and a harmonic spring potential to model the bead-spring interactions such that

\begin{align}
u_{ij}^B &= \frac{a_{ij}}{2r_c}(r_{ij}-r_c)^2,
\label{eq:bbPotential}
\\
u_{ij}^S &= \frac{K_s}{2}(r_{ij}-r_o)^2, 
\label{eq:bsPotential}
\end{align}
where $a_{ij}$ is an interaction (repulsion) parameter, $K_s$ is the spring constant and $r_o$ corresponds to the equilibrium average distance between particles, defined in \eqref{eq:meansquaredistance}.
\\
\\
If the system undergoes a first order transition under compression, the change in the distance between particles can be written as $\Delta r_{ij}(\zeta) = \zeta r_{ij} + \mathcal{O}({\zeta^2})$. If we substitute \eqref{eq:bbPotential} in \eqref{eq:scalebyrc}, taking a first order approximation  yields 
\begin{align}
\bar{a}_{ij} &= \frac{\frac{1}{{r'}_c}(({r'}_{ij} - \zeta {r'}_{ij} - {r'}_c)^2 - ({r'}_{ij}-{r'}_c)^2)}{\frac{1}{\bar{r}_c}((\bar{r}_{ij} - \zeta \bar{r}_{ij} - \bar{r}_c)^2 - (\bar{r}_{ij}-\bar{r}_c)^2)} a'_{ij},\nonumber
\\
&= \frac{\frac{1}{{r'}_c}({r'}_{ij}^2\zeta^2 - 2\zeta {r'}_{ij}^2 + 2{r'}_{ij}{r'}_c\zeta)}{\frac{1}{\bar{r}_c}(\bar{r}_{ij}^2\zeta^2 - 2\zeta \bar{r}_{ij}^2  +2\bar{r}_{ij}\bar{r}_c\zeta)} a'_{ij}, \nonumber
\\
 &= \frac{(1-\frac{{r'}_{ij}}{{r'}_c}){r'}_{ij}}{(1-\frac{\bar{r}_{ij}}{\bar{r}_c}) \bar{r}_{ij}} a'_{ij}. \label{eq:scalebyrcExplicit}
\end{align}
Since the cutoff radius and the distance between particles have the same order of magnitude, the ratio $\frac{r_{ij}}{r_c}$ between fine and coarse scales is conserved, and the remaining terms of equation \eqref{eq:scalebyrcExplicit} that contribute to the scaling are 
\begin{align}
\bar{a}_{ij} &= \frac{{r'}_{ij}}{\bar{r}_{ij}} a'_{ij}.
\label{eq:scalebyrcFinal}
\end{align}
From equation \eqref{eq:rcscaling} we know that the length ratio between scales is $\frac{{r'}_c}{\bar{r}_{c}} = N'^{-\nu' + \bar{\nu}}\phi^{-\bar{\nu}}$. Combining equations \eqref{eq:scalebyrcFinal} and the scaling originated by the change in number of particles \eqref{eq:scalebyN} $\bar{a}_{ij} = \phi a'_{ij}$, the scaling of the conservative contribution is finally obtained as
\begin{equation}
\bar{a}_{ij} =N'^{-\nu' + \bar{\nu}}\phi^{1-\bar{\nu}} a'_{ij}.
\label{eq:aijfromU}
\end{equation}
Once the scaling for the bead-bead interactions is identified, the remaining conservative contribution to scale is the bead-spring potential used to construct DPD chains. From equation \eqref{eq:scalebyN} and substituting \eqref{eq:bsPotential} in \eqref{eq:scalebyrc} the same procedure used to scale the interaction parameter is applied. In this case, the scaling due to the change in length scale is $\phi^{-2\bar{\nu}} N'^{(2\bar{\nu}-\nu')}$. While the scaling of bead-spring potentials due to the change in the particle density, leads to the  scaling $(\phi + (\phi-1)\phi/{N'})$. Where the density scaling given in equation \eqref{eq:scalebyN} takes into account the change in particle density.The resultant spring constant in the coarse-grained representations is		

\begin{equation}
\bar{K_s} = (N' +\phi-1)\phi^{1-2\bar{\nu}}N'^{2(\nu'-\bar{\nu})-1} K'_s
\label{eq:ksScaling}
\end{equation}

\subsubsection*{Remark}
 An alternative scaling based on the analysis of the conservative forces can be used following the approach proposed by Backer et. al.\cite{Backer2005} Nevertheless, we have verified that the methodology proposed in \cite{Backer2005} and \cite{Fuchslin2009} are equivalent. We now derive the scaling of the interaction parameter $a_{ij}$ following the procedure proposed in \cite{Backer2005} by the authors. In this case, we seek to preserve the pressure of the system. The pressure can be expressed using the virial theorem,\cite{Groot1997} and written as a summation over the particles in the system

\begin{equation}
p = \hat{\rho} k_BT + \frac{1}{3V}\ave{\sum_{j>i} (\textbf{r}_i - \textbf{r}_j)\cdot \textbf{F}_{ij}^C},
\label{eq:pressSum}
\end{equation}
where $\hat{\rho} = N_T/V$ is the particle number density, while
$N_T$ and $V$ are the total number of particles and volume, respectively. $\textbf{F}_{ij}^C$ is the conserved part of the force on the
particle $i$. Equation \eqref{eq:pressSum} is valid since the
dissipative and random forces have been defined to be a Boltzmann
distribution. \cite{Search1995} In \eqref{eq:pressSum}, the first term of the right-hand side accounts for the ideal contribution to the pressure while the second one accounts for the residual contributions.
\\
\\
We start out by analysing the pressure preservation at the ideal condition where $a_{ij} = 0$. In this case we expect $p|_{a'=0} = p|_{\bar{a}=0}$. According to our selection for the energy scaling
\eqref{eq:energyScaling}, $\bar{\epsilon} = \phi \epsilon'$ and $\bar{\rho} =
{\bar{N}_T}/{\bar{V}} = {\phi^{-1} N'_T}/{V'}$, thus we have $\bar{\hat{\rho}}\overline{\epsilon} = \hat{\rho}' \epsilon'$, which yields     
\begin{equation}
\bar{\hat{\rho}}(\overline{k_BT}) = \hat{\rho}' (k_BT)'.  
\label{eq:idealp}
\end{equation}
Since the ideal contribution of the pressure is independent of the coarse graining, the scaling of the conservative bead-bead interaction can be expressed as 

\begin{equation}
\bar{a}_{ij} = \psi_c (\phi) a_{ij}',
\label{eq:aijmoreno}
\end{equation}
where the interaction scaling $\psi_c (\phi)$ is only related to the residual term of the pressure, which from \eqref{eq:pressSum} requires that,

\begin{equation}
\frac{1}{3V'}{\sum_{i=1}^{N'}\sum_{j>i}^{N'} (\textbf{r}_i - \textbf{r}_j)\cdot {\textbf{F}'}_{ij}^C} = \frac{1}{3\bar{V}}{\sum_{i=1}^{\bar{N}}\sum_{j>i}^{\bar{N}} (\textbf{r}_i - \textbf{r}_j)\cdot \bar{\textbf{F}}_{ij}^C}.
\label{eq:scaleForce}
\end{equation}
If the bead-bead potential is given by equation \eqref{eq:bbPotential}, the analysis performed in equation \eqref{eq:scalePotential} can be directly used to \eqref{eq:scaleForce}, leading to an equivalent scaling of equation \eqref{eq:aijfromU}. Therefore, we conclude that we have the same scaling of $a_{ij}$ due to the change in the particle density and length scale. 
\\
\\
In the methodology proposed by Backer\cite{Backer2005} and Spaeth,\cite{Spaeth2011} due to the absence of an explicit energy scaling, $(k_BT)' = (\overline{k_BT})$, and thus they found that the ideal pressure depends on the coarse graining ($p|_{a'=0} \neq p|_{\bar{a}=0}$). This dependency requires the scaling of $a_{ij}$ to include a correction term $\psi_o(\phi)$ for the ideal pressure, such that 

\begin{equation}
\bar{a}_{ij} = \psi_o(\phi) + \psi_c (\phi) a_{ij}'.
\label{eq:aijspaeth}
\end{equation}
Equation \eqref{eq:aijspaeth} effectively produces $\bar{a}_{ij} \neq 0$ when $a_{ij}'=0$. The first term on the right-hand side of \eqref{eq:aijspaeth} accounts for the ideal pressure, therefore, likewise equation \eqref{eq:aijmoreno},  $\psi_c (\phi)$ only needs to ensure the preservation of the residual pressure contribution. According to our definition of coarse grain level ($\phi$), and the scaling of $a_{ij}$ presented in, \cite{Spaeth2011} we identify that in equation \eqref{eq:aijspaeth}  $\psi_{c}(\phi) = \phi^{2/3}$, which results to be equivalent to the scaling proposed by Fuchslin et.al.,\cite{Fuchslin2009} Hence, the only difference between the methodologies followed by \cite{Backer2005} and \cite{Fuchslin2009} is the correction of the ideal pressure proposed  in \cite{Backer2005} due to the absence of a consistent energy scaling.

\subsection{Scaling for dissipation $\gamma$ and
  fluctuation $\sigma$ parameters}

We now consider the scaling of the friction $\gamma$ and noise
$\sigma$ coefficients. Taking into account the scaling we already chose for time and energy units, we scale the friction and noise parameters as 

\begin{align}
\bar{\gamma} &= \phi^{1-\bar{\nu}} N'^{(\bar{\nu}-\nu')}\gamma',
\\
\bar{\sigma} &=  \phi^{1-\frac{\bar{\nu}}{2}} N'^{\frac{(\bar{\nu}-\nu')}{2}} \sigma'.
\end{align}
 
In this case, the methodologies proposed in \cite{Fuchslin2009} and \cite{Backer2005} lead to different scaled parameters. In \cite{Backer2005} the authors attempted to scale the parameters to preserve the viscosity of the system, while \cite{Fuchslin2009} scales these parameters based on dimensional analysis. Thus, the scaling of the dissipation and fluctuation parameters that we use, share the same foundations of those proposed by Fuchslin et al.\cite{Fuchslin2009}  

\subsection{Coarse-chain conformation}

\begin{figure}[h!]
\centering
\includegraphics[width=0.75\linewidth]{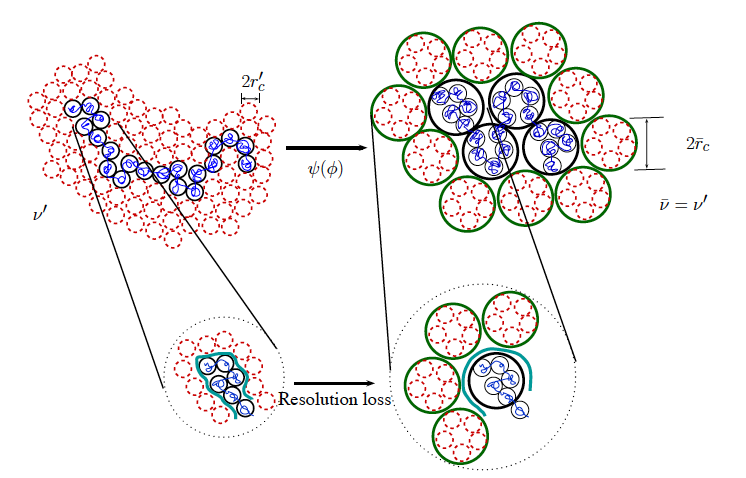}
\caption{Schematic representation of the coarse graining process. In this case sets of five fine particles are grouped into single coarse particles representations. The current parameter scalings preserve the number of interactions per particle, but does not explicitly account for the change in the number of polymer-solvent interactions. Therefore long-range correlations vanish. The grouping of fine particles originates a reduction in the accessible area of a chain due to the change in the contour length $l_c$.}
\label{fig:coarseGraining}
\end{figure}

\begin{figure}[h!]
\centering
\includegraphics[width=0.75\linewidth]{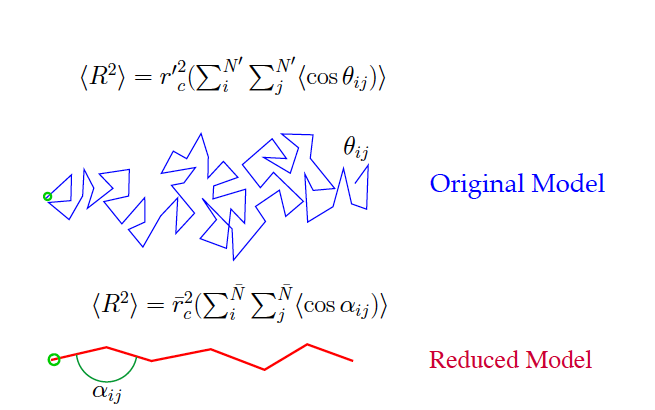}
\caption{Bond-angle restriction proposed for coarse chain models to satisfy $\bar{\nu}$ such that the size of the polymer is preserved. }
\label{fig:angleModelReduction}

\end{figure}

The choice of scaling $\phi$ imposes restrictions on the coarse chain configuration; in particular, the coarse chain must satisfies $R= \bar{N}^{\bar{\nu}}\bar{r}_{c}$. Abstractly, one controls the configuration of the coarse chain by modulating either the entropic or the enthalpic interactions. In this paper we control the coarse chain conformation entropically. 
\\
\\
As shown in Figure \ref{fig:coarseGraining}, the coarse graining of a chain reduces its contour length, and so the area of the chain that is accessible to the solvent is reduced. We denote this effect \textit{resolution loss}, which affects the effective contact between the polymer and the solvent. The proposed scaling of the interaction parameter $a_{ij}$ preserves the enthalpic contribution accounting for the change in the total number of interactions in the system (i.e., particle density) and the length scale of these interactions (i.e., cutoff radius), such that the number of interactions per particle  ($g(r/{r'}_c) = g(r/\bar{r})$) is assumed independent of the coarse graining level. However the resolution loss also affects how the solvent particles localize around the polymer chain.  The localization of the solvent defines the number of polymer-solvent interactions, and originates the long-range correlations between segments. Since the current scaling of the conservative contributions does not account for this solvent localization, coarse-grained system would tend to reproduce the same chain conformation for any $\phi$. 
\\
\\
In order to identify how the required coarse conformation $\bar{\nu}$ can be achieved, we analyse the effect of the resolution loss considering the free energy of the DPD chain. The free energy $\mathcal{F}$ of a fine-grained chain can be defined as $\mathcal{F'} = \mathcal{F'}_{h} + \mathcal{F'}_{ent}$,\cite{Rubinstein2003} where the subscript \textit{h} and \textit{en} denote the enthalpic and entropic components of the free energy, respectively.
\\
\\
The enthalpic contribution of the free energy is obtained if we define the probability to find a chain segment within the cutoff radius of another segment, as the product of the bead volume (${r'}_c^3$) and the number density of segment inside the pervaded volume of the chain ($N'/{r'}^3$). Such that we can express the enthalpic interaction per segment as $\epsilon'({r'}_c^3N'/{r'}^3)$, where $\epsilon'$ indicate energy units. The interaction energy for the whole chain is written as
\begin{equation}
\mathcal{F'}_{h} =  \epsilon'{r'}_c^3 \frac{N'^2}{{r'}^3} =  \epsilon' \frac{N'^2}{N'^{3\nu'}}.
\end{equation}
If we now write the interaction energy of a coarse-grained chain with the same conformation, we obtain that
\begin{equation}
\mathcal{\bar{F}}_{h}  =  \overline{\epsilon} \bar{r}_c^3 \frac{\bar{N}^2}{\bar{r}_c^3 \bar{N}^{3\nu'}}=\mathcal{F}'_{h} \phi^{3\nu'-1}.
\label{eq:freeEint}
\end{equation}
From equation \eqref{eq:freeEint} we find that $\mathcal{\bar{F}}_{h}\geq \mathcal{F}'_{h} $. Therefore, if the interaction energy needs to be preserved we require for the coarse chains that

\begin{equation}
\mathcal{\bar{F}}_{h}  = \mathcal{F}'_{h} \phi^{3\nu'-1} + \psi_{h}(\phi),
\end{equation}
where $\psi_{h}(\phi)$ is a correction term that compensates the change in $\mathcal{F}_{h}$ due to the model reduction.
\\
\\
Similarly to the Flory theory,\cite{Rubinstein2003} we now estimate the entropic contribution to be the energy required to deform the fine chain from it theta condition to the current fine-chain dimension, leading to 

\begin{equation}
\mathcal{F'}_{ent} =  \epsilon'\frac{R^2}{R_{\theta}^2} =  \epsilon'\frac{{r'}_c^2N^{2\nu}}{{r'}_c^2N} =  \epsilon' N'^{(2\nu'-1)}.
\end{equation}
While for the coarse-grained counterpart the entropic contribution to free energy is
\begin{equation}
\mathcal{\bar{F}}_{ent} = \overline{\epsilon} \bar{N}^{2\nu'-1} = \epsilon'\phi \left(\frac{N'}{\phi}\right)^{2\nu'-1} = \mathcal{F}'_{ent} \phi^{2-2\nu'}. 
\end{equation}
As for the interaction energy, we identify that $\mathcal{\bar{F}}_{ent}\geq\mathcal{F}'_{ent} $, which in turn requires a correction factor $\psi_{ent}(\phi)$ to the entropic term of the coarse scale system, such that
 
\begin{equation}
\mathcal{\bar{F}}_{ent} = \mathcal{F}'_{ent} \phi^{2(1-\nu')} + \psi_{ent}(\phi).
\end{equation}
According to the definition of the free energy for the fine and coarse chains, we find that the model reduction process originates a resolution-loss effect that requires and additional energetic term contribution that fixes both enthalpic and entropic components of the free energy. To address this shortcoming, we use bond-angle potentials that allow us to achieve an effective $\bar{\nu}$ for a given coarse graining level. In the model-reduction methodology proposed this additional potential can be interpreted as the required correction to the free energy of the chain. The bond-angle potential we use is given by

\begin{equation}
u_{ij}^A = \frac{\bar{K}_a}{2}(\alpha_{ij} - \alpha_o)^2,
\end{equation}
where $K_a$ is the bending constant, $\alpha_o$  is the equilibrium
magnitude of the angle and $\alpha_{ij}$ is the current angle between the
bond vectors $\textbf{s}_i$ and $\textbf{s}_j$. The bond-angle potentials control entropically the coarse-chain conformation and account for long-range correlations eliminated by the coarse-graining process as schematically shown in Figure \ref{fig:angleModelReduction}. The inset equations in Figure \ref{fig:angleModelReduction} show that coarse graining affects the length scale and the number of terms considered in the definition of mean-square radius for the fine and coarse models. Thus,  when the bond-angle potential is imposed we are effectively scaling the proper cosine similarity average.
\\
\\

\begin{figure}[h!]
\centering
\includegraphics[width=0.9\linewidth]{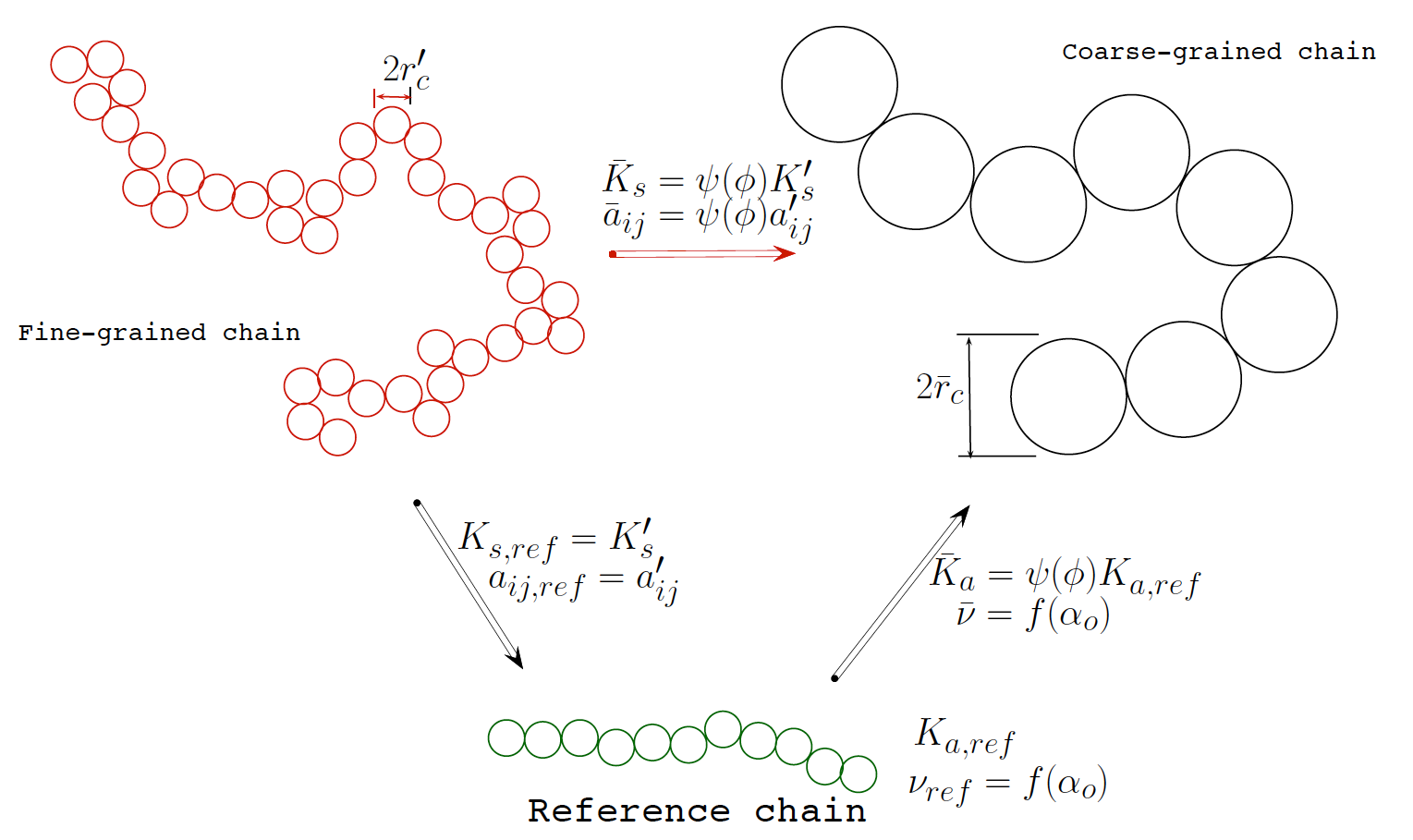}
\caption{Identification of the scaling of entropic restriction through a reference chain}
\label{fig:finerefcoarse}

\end{figure}

\noindent
Since the main function of the bond-angle potentials is to achieve a required $\bar{\nu}$, we could fine tune the bending constant $\bar{K}_a$, and the equilibrium angle $\alpha_o$ for a given coarse grain level. However this approach requires an iterative process every time we change the coarse-graining level.  To avoid this issue we construct a \textit{reference} chain with $\phi_{ref}=1$ and bond-angle potentials, where we perform the tuning of $\bar{K}_a$ and  $\alpha_o$ that compensates the bead-bead and bead-spring interactions such that the chain conformation is entropically controlled. Based on this reference chain we identify the magnitude of the bending constant $\bar{K}_a$ that does not distort the pressure of the system, as well as the relationship

\begin{equation}
\bar{\nu} = f(\alpha_o).
\label{eq:nufofAlpha}
\end{equation}
Thus, for an arbitrary coarse-graining level we can scale the bending constant from this reference chain and the conformation of the coarse-grained chain can be controlled through the expression \eqref{eq:nufofAlpha}. Here, we assume that \eqref{eq:nufofAlpha} is independent of the model reduction process, which implies that there is no change in the angle length scale  
\begin{equation}
\Delta \alpha_{ij,ref} = \Delta \bar{\alpha}_{ij}.
\label{eq:alphaNoCh}
\end{equation}
To simplify the process, we construct the reference system using the same bead-bead and bead-spring interactions of the fine-scale system. Figure \ref{fig:finerefcoarse} illustrates the methodology that we adopt to scale the bond-angle restrictions.

The proper scaling of the bond-angle constant is determined from equation \eqref{eq:scaleConservative}, stressing the same arguments we use to scale the bead-bead and bead-spring interactions. In this case the bending constant only scales with the change in the number of angles ($\bar{N}_{angles} + 2 = (N'_{angles} + 2)\phi^{-1}$), since the length scale of the angle is unchanged \eqref{eq:alphaNoCh}. The scaling of the bending interactions yields 

\begin{equation}
\bar{K}_a = (N'+\phi-2)\frac{\phi}{N'}K_{a,ref},
\label{eq:scalingKa}
\end{equation}
where the subscript $ref$ indicates that this constant is applied over a reference system. In Section \ref{sec:validation} we discuss further the construction of the reference chain.
\\
\\ 
We summarize the scaling functions proposed herein in Table \ref{tab:scalings}. We stress the fact that the two key features of this methodology are the explicit consideration of the chain conformation ($\nu'$ and $\bar{\nu}$), and the bond-angle potential that allows us to control the conformation of the coarse scales.

\begin{table}
 \caption {Coarse parameters $A_{coarse}$ and scaling function $\psi(\phi)$ proposed for coarse graining of systems with chains. }
 \centering
    \begin{tabular}{c|c}
    \hline
   \textbf{$A_{coarse}$} \quad   &  \textbf{$\psi(\phi)$}     \quad      \\ \hline \hline                                                                                                                              
 $\bar{m}$                 & $\phi$                                  \\ \hline                                            
 $\bar{N_T}$             & $\phi^{-1}$                          \\ \hline                                               
 $\bar{r}_c $               & $N'^{\nu' - \bar{\nu}}\phi^{\bar{\nu}}$     \\ \hline                                 
 $\bar{v}    $               & $(N'^{(\nu' - \bar{\nu})}\phi^{\bar{\nu}})^3$         \\ \hline                              
 $\bar{\nu}  $             & $\frac{1}{3}~\frac{({3\nu' \text{ln}N' - \text{ln}\phi})}{({\text{ln}N' - \text{ln}\phi})}$ \\ \hline
 $\bar{\tau}  $           &  $N'^{(\nu' - \bar{\nu})}\phi^{\bar{\nu}}$      \\ \hline                                     
 $\bar{\epsilon}$        & $\phi$                                                        \\ \hline                      
 $\bar{a}_{ij} $             & $\phi^{1-\bar{\nu}} N'^{(\bar{\nu}-\nu')}$      \\ \hline
 $\bar{K_s} $                & $(N' +\phi-1)\phi^{1-2\bar{\nu}}N'^{2(\nu'-\bar{\nu})-1}$ \\ \hline 
 $\bar{\gamma}$         & $\phi^{1-\bar{\nu}} N'^{(\bar{\nu}-\nu')}$                                                        \\ \hline 
 $\bar{\sigma}  $          & $\phi^{1-\frac{\bar{\nu}}{2}} N'^{\frac{(\bar{\nu}-\nu')}{2}}$     \\ \hline
 $\bar{K_a} $                 &$(N'+\phi-2)\frac{\phi}{N'}$    \\ \hline
    \end{tabular}
 \label{tab:scalings}
\end{table}
 
\section{Model Reduction Validation}
\label{sec:validation}

\subsection{Simulation Details}
\label{subsec:simdetails}

\begin{figure}[h!] 
\begin{center}
\includegraphics[width=0.75\linewidth]{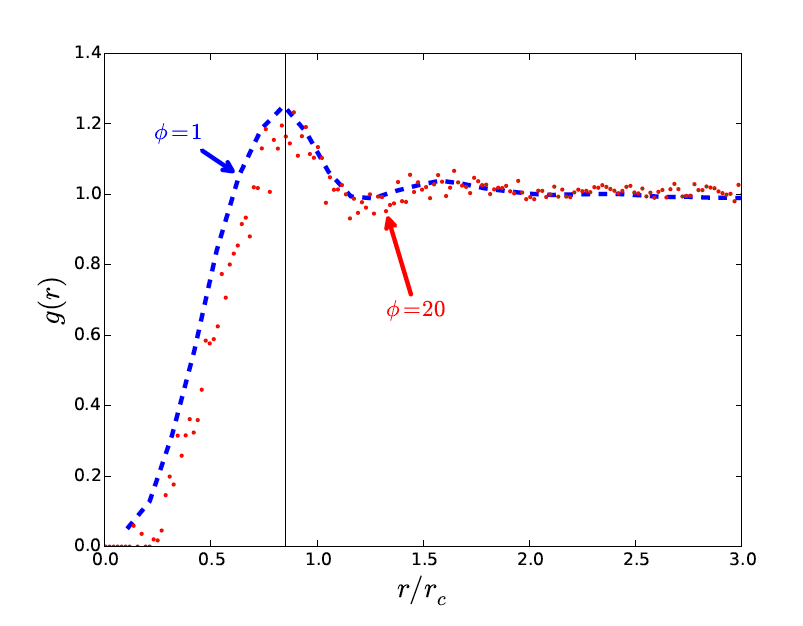}
\end{center}
\caption{Comparison of the radial distribution function for a fine scale system and it coarse-grained counterpart. The maximum peak occurs at $r/r_c \approx 0.85$.}
\label{fig:gofr}
\end{figure}

The DPD simulations are conducted using the software LAMMPS.\cite{Plimpton1995} The simulation box size for the systems modeled ranges from $(30r_c)^3$ to $(100r_c)^3$, with chain lengths ranging from $4$ to $400$ beads. For the fine-grained systems we fix the scales of energy $\epsilon' = k_BT$, length ${r'}_c = 1$, and mass $m'=1$. This choice leads to the standard (see equation \eqref{eq:tau}) time unit $\tau' = 1$. In those fine systems the particle density used is $\hat{\rho}' = 3$ particles$/r_c^3$.  We adopt a time step $\Delta \tau = 0.04$ in order to have temperature fluctuations smaller than $2\%$. The simulations run for $500,000$ time steps, including an initial stabilization period of $20,000$ time steps. During the stabilization stage, the interaction parameters are selected to be $a'_{ps} = 25$, where $p$ and $s$ denote polymer and solvent, respectively.  The units used for each coarse-grained simulation are set according to the framework described in section \ref{sec:coarsegraining} and summarized in Table \ref{tab:scalings} 
\\
\\
The radius of gyration, end-to-end distance, and contour length are averaged for production runs of $500.000$ steps, sampling every $1.000$ steps. In order to get significant statistical results in the measurement of the exponent $\nu$, the sampling frequency is increased to every $500$ steps.
\\
\\
In all fine-scale simulations, the size of the simulation boxes is chosen proportional to the expected end-to-end distance as $2R_{f,expect}$. We define

\begin{equation}
R_{f,expect} = r_o(N')^{\nu_{e}},
\label{eq:rfexpected}
\end{equation}  
where $\nu_e=0.65$ for interaction parameters polymer-solvent ($a_{ps}$) smaller than $29$, and $\nu_e=0.4$ otherwise. According to our experience and the literature  \cite{Yang2013, Ilnytskyi2007} the theta condition occurs at $27 \leq a_{ps} \leq 28$, therefore we expect for  $a_{ps}>29$ that the DPD chains adopt collapsed configurations. Similarly, below the theta condition extended structures are preferred, and larger boxes  are required to avoid finite size effects.
\\
\\
Similarly to Spaeth et al.,\cite{Spaeth2011} we select the equilibrium distance between connected particles $r_o$ to be equal to the distance where the maximum of the radial distribution function $g(r)$ occurs. To corroborate our selection of $r_o$,  the $g(r)$ of the particles in the fine and coarse systems were evaluated. Our estimates coincide with the results of Spaeth,\cite{Spaeth2011} leading $r_o \approx 0.85 r_c$. In Figure \ref{fig:gofr} the radial distribution function measured is presented for systems containing fine grained chains and coarse graining representations with $\phi=20$. We track the variation of the particle distance between connected particles to to verify that $r_o$ is satisfied. The chain size calculations are computed from the measured contour length, such that the measured average is given by $r_o^{	av} = \bar{l}_c/\bar{N}$  . 

In all the systems the Flory-like ratio parameter $C_N$ was measured for different chains and the characteristic $\nu$ was determined  as

\begin{equation}
\nu = \frac{1+\beta}{2},
\end{equation}
where $\beta = {\text{ln}C_N}/{\text{ln}N'}$ from equation \eqref{eq:floryratio}.
The influence of the chain length and the solvent interactions on the bead correlation along the chain is studied by computing the bond-angle correlation function $B(n)$, defined for a chain with $N$ beads as

\begin{equation}
B(n) = \frac{1}{N-n+1}\sum_{m=1}^N C_{nm},
\label{eq:bondAngleCorrelation}
\end{equation}  
where $C_{nm}$ is the Flory's characteristic ratio of chain fragments containing $n$ segments, and is given by

\begin{equation}
C_{nm} = \frac{1}{n}\sum_{i=m}^{n+m-1}\sum_{j=m}^{n+m-1} \ave{\cos \theta_{ij}}.
\end{equation}
The expression \eqref{eq:bondAngleCorrelation} accounts for all the possible fragments of size $n$ in the chain. Equation \eqref{eq:bondAngleCorrelation} is useful to compute the change of the conformation ($\nu$) between segments along a DPD chain. Thus, we can identify long range correlations and their relationship with the chain length and solvent interaction.

\subsection{Reference chain construction}

\begin{figure}[h!] 
\begin{center}
\includegraphics[width=0.75\linewidth]{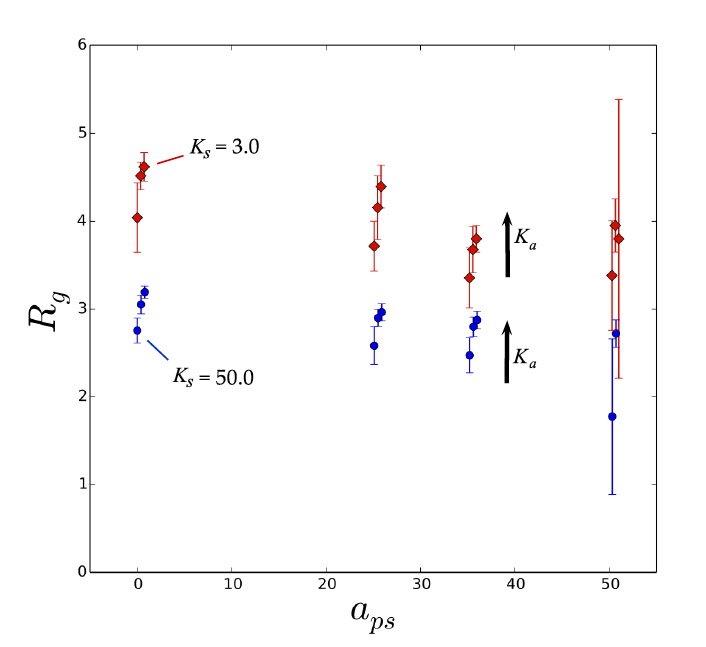}
\end{center}
\caption{Radius of gyration of a single DPD chain for different interaction parameter $0.0 \leq a_{ps} \leq 50$. The chains are constructed with spring constants $K_s=3.0$ and $K_s=50.0$. For each set of $K_s$ the arrows indicate the effect of increasing in bending constant. The error bars are included to indicate the standard deviation, and the data points have been shifted horizontally to facilitate the visualization of the error bars.}
\label{fig:kseffectOnRg}
\end{figure}

We conduct the inter- and intra-chain interaction experiments in a reference grained system basis. Thus, the appropriate spring and bending parameters are identified for reference chain and scaled consistently to be used in the coarse-grained scales. 
\\
\\
In order to satisfy the proper correlation $\bar{\nu}$ between particle chains in entropically constrained models, we study how the bending constant $K_{a,ref}$ and the equilibrium angle $\alpha_o$ influence the reference-chain conformation. The intervals evaluated for these parameters are listed in the table \ref{tab:shapeDetails}.  In the simulations we vary independently $K_{a,ref}$ and $\alpha_o$, for different interaction parameters $a_{ps}$ and spring constants {$K_{s}$}.  The goal is to assess the effect of the bending constant ($K_{a,ref}$) on the balance between intra- and inter-molecular forces acting on a given particle, and the effect of the equilibrium angle in the magnitude of the Flory's ratio (and consequently $\nu$).  In this case the magnitude of the spring constant chosen is used to construct all the fine-scale systems.
\begin{table}
 \caption {Parameters and ranges evaluated for the entropic-constraint studies in reference chain models.}
  \centering
    \begin{tabular}{lr}
    \hline
    Parameter & Values                                                                    \\ \hline
    $K_{s}$      & 3 -  50                                                         \\ \hline
    $K_{a,ref}$      & 0 - 10                                                    \\ \hline
    $a_{ps}$     & 0 - 60  \\ \hline
    $N$         & 4, 8, 10, 12, 16, 32                                                      \\ \hline
    $\alpha_o$ & 90 - 180                                          \\ \hline
    \end{tabular}

 \label{tab:shapeDetails}
\end{table}
\\
\\
We identify the interval in which the characteristic size of the polymer chain can be properly controlled by tuning the entropic restrictions.  Furthermore, we ensure that the reference chain conformation can be driven by the bond-angle potential, without affecting on the enthalpic interactions.  In Figure \ref{fig:kseffectOnRg} we present the radius of gyration of  a DPD chain constructed with the lower and upper limit of spring constants evaluated $K_{s}=3.0$ and $K_{s}=50.0$, respectively. For both values, the variation in size with the bending constant $K_{a,ref}$, for fixed equilibrium angle, $\alpha = 180^o$, is shown.   In Figure \ref{fig:kseffectOnRg} the highest value of the bending constant corresponds to $K_{a,ref} = 9$, while the lowest $K_{a,ref}=3$.
\\
\\
It is identified in Figure \ref{fig:kseffectOnRg} that the chain size controllability is diminished at low values of $K_{s}$; in this condition, the strength of the enthalpic interaction between the chain and the solvent induces large fluctuations in the average bond distance $r_o^{av}$, increasing the variance on the chain size dimensions. In contrast, when bond interactions are stronger the bead-bead contributions are damped, and the entropic restriction dominates. In addition  from Figure \ref{fig:kseffectOnRg}, it is evident that the better entropically-governed chain models occur when the bending constant takes the maximum value in the interval evaluated.  In general, we found that the highest values of $K_{s}$ and $K_{a,ref}$ have the best performance to entropically control the size of the DPD-reference chains, in a narrow fashion. 
\\
\\
The identification of the inter- and intra-molecular interactions is analyzed by studying their effect on the chain conformation $\nu$. Similarly to the the effect on radius of gyration,  higher values of $K_s$ and $K_a$ improves the control of the chain conformation.  Based  on these observations we select the bead-spring ($K'_s = 50 k_BT$)  and bond-angle ($K'_a = 9 k_BT$) interactions of the reference chain that consistently preserve $\nu$ for different conditions of polymer length and solvent interaction. For longer chains (i.e., 32 beads per chain) it is necessary to use higher values of the bending constant $K'_a>10$. For long chain models folding of the structure is possible, conserving local rigidity in a short sequence of beads. In the case of a polymer chain with $32$ beads the number of rigid sections identified is $\approx4$.  

\subsection{Identification of the coarse-scale conformation $\bar{\nu}$}
\label{subsec:coarsenu}
\begin{figure}[h!] 
\begin{center}
\includegraphics[width=0.75\linewidth]{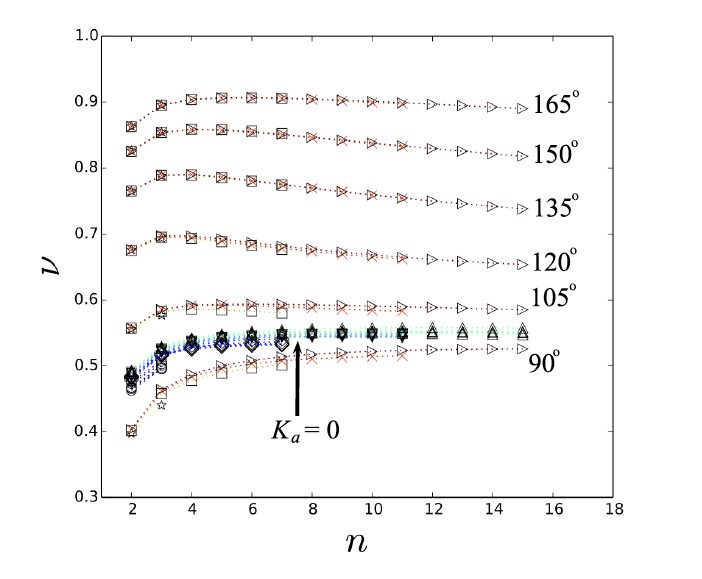}
\end{center}
\caption{Variation of particle correlation $\nu$ for different equilibrium angles $\alpha$ in the bond-angle potentials applied to entropically constrained chain models, in the athermal condition, $a_{ps}=25.0$. For the different values of $\alpha$ presented, we included as a comparison the particle correlation of chain models without bond-angle potentials. The variation of $\nu$ is presented along chains with $N$ ranging from 4 to 16.} 
\label{fig:angleVari}
\end{figure}

\begin{figure}[h!]
\begin{center}
\includegraphics[width=0.75\linewidth]{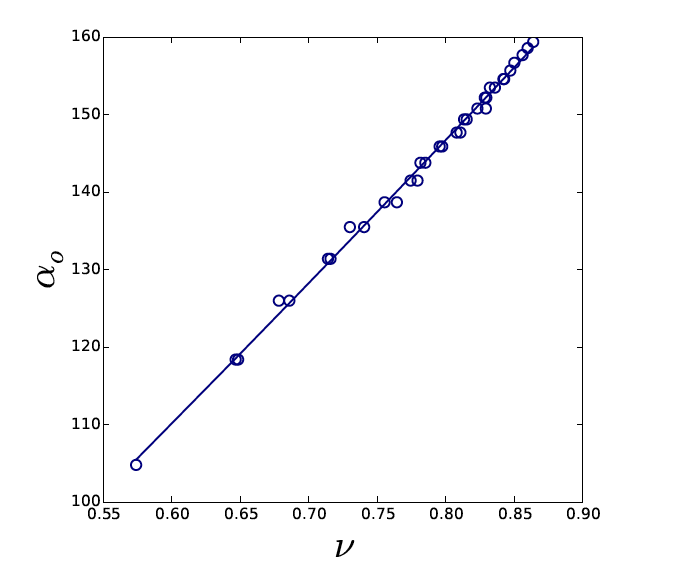}
\end{center}
\caption{Equilibrium angle $\alpha_o$ variation with the average segment conformation $\nu$. The averages values are computed for DPD-chain models ranging from $4$ to $16$ beads per chain. The size of the symbols correspond to the standard deviation of the data}
\label{fig:nuwithangle}
\end{figure}

The variation of the chain conformation $\nu$ with the imposed angle is measured for reference chains. The magnitude of the angles and polymer lengths evaluated in the reference systems are presented in the Table \ref{tab:shapeDetails}. 
\\
\\
The effect of the equilibrium angle $\alpha$ in the bond-angle correlation function $B(n)$ (defined in equation \eqref{eq:bondAngleCorrelation}) of athermal systems ($a_{ij} = a_{ii}$), is presented in Figure \ref{fig:angleVari}. In this case we analyse short chain models to show how the coarse-graining process proposed is able to accurately represent long chains using short-coarse models. Figure \ref{fig:angleVari} exhibits the variation of $\nu$ for chains containing 4 to 16 beads per chain, evidencing that we can narrowly control the value of $\nu$. As a comparison, the correlation $\nu$ between particles for chain models  without bond-angle potential is also included. We found that in these short-chain models due to the absence of long-range correlations, the variation of $\nu$ with the molecular weight is negligible, and the bond-angle correlation function for all the chains collapses in a single curve. The angle imposition uniformly modifies the correlations between segments within the chains, preserving the molecular weight independence.  
\\
\\
In Figure \ref{fig:nuwithangle} we summarize the average $\nu$ variation with the equilibrium angle imposed, the variation of $\nu$ and $\alpha$ is approximated by a linear function, such that we can compute the angle to achieve the sought $\bar{\nu}$. Thus, we express the required angle as

\begin{equation}
\alpha_{ij} =  -1.55 + 185.56\nu,
\label{eq:alphaFunction}
\end{equation}
with a standard deviation $\pm 0.02$, and $\nu$ in the range $1/3$ to $1$.

\subsection{Identification of fine-scale conformation $\nu'$}
\label{subsec:finenu}
\begin{figure}[h!] 
\begin{center}
\includegraphics[width=0.75\linewidth]{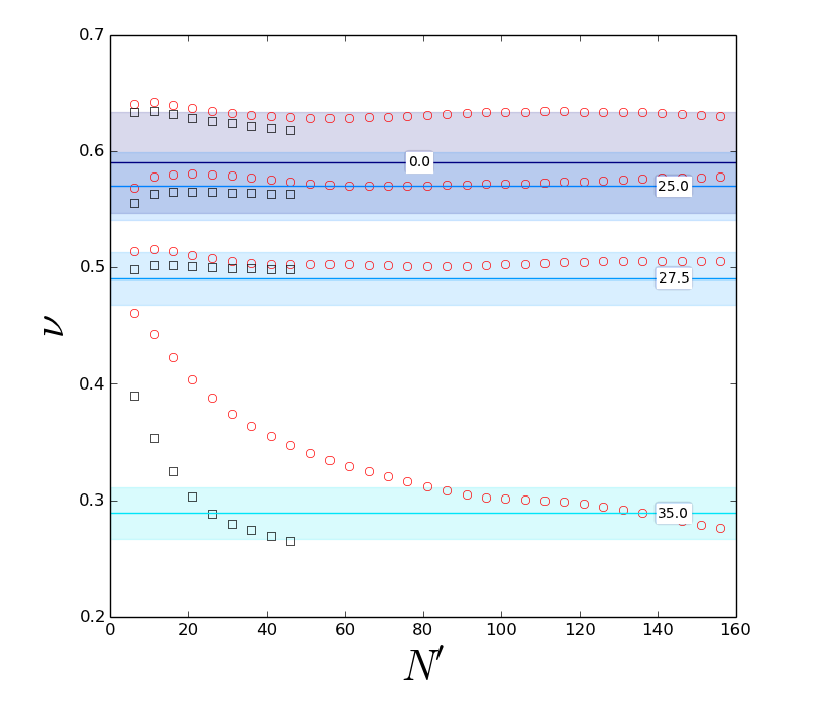}
\end{center}
\caption{Effect of $a'_{ij}$ over the particle correlation along the chain models with 50 and 160 beads, with fixed $K'_s = 50 k_BT$, and  $K'_a = 0.0 k_BT$. The average $\nu'$ computed for chains ranging from $50$ to $400$ beads is depicted a solid lines for different $a'_{ij}$. The shadowed regions indicate the standard deviation of the $\nu$ computed.}
\label{fig:aijEffect}
\end{figure}

Due to the relevance of the fine grained conformation $\nu'$ in the coarse-graining methodology proposed, we study the behavior of $\nu'$ in a DPD system as the interactions between components varies. Polymer chains ranging from 16 to 400 beads per chain were modelled under different solvent affinity conditions. In order to preserve the chain controllability for different levels of coarse graining, the magnitude of the spring constant identified for the reference chains is used for all the fine systems evaluated.  To ensure the chain incompressibility condition in fine-scale systems, we identify the interval of the interaction parameter $a'_{ij}$ that satisfies $0.33 \leq \nu' \leq 1.0$. 
\\
\\
In Figure \ref{fig:aijEffect} we  use the bond-angle correlation function given in \eqref{eq:bondAngleCorrelation} to compute the variation of $\nu'$ along chains containing $50$, $200$ and $400$ beads, for different $a_{ps}$.  As we already mention, the limits on the values of $\nu'$ can be interpreted from a geometrical standpoint, however this interval holds only under incompressibility constraints. We observed that for the magnitude of the spring constants chosen ($K'_s=50$), the maximum value of interaction the parameter that does not induce significant compression in the chain is $0<a_{ps}<35$, if $N'$ is sufficiently large (i.e., $N'>100$ beads). In addition, the analysis of $B(n)$ over different polymer lengths reveals that the incompressibility is satisfied not only at the chain-scale level but also locally along the chain. Based on this result we identify  $a_{ps}=35$ as the non-solvent limit for the systems evaluated. In general, a system-specific mapping may require higher values of $a_{ps}$ which in turn would need a compensation with $K'_s>50$ to satisfy incompressibility, or the use of different bead-spring potentials. 
\\
\\

\begin{figure}[h!] 
\begin{center}
\includegraphics[width=0.75\linewidth]{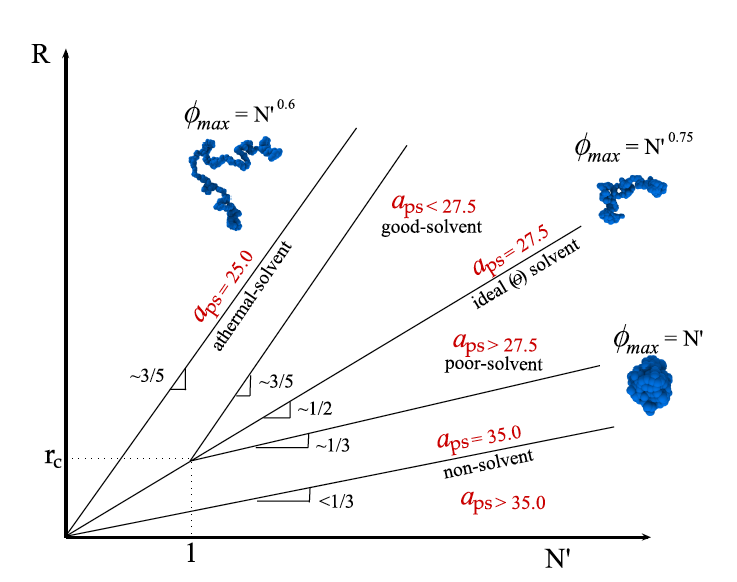}
\end{center}
\caption{Variation of $R_g$ for different interaction parameters (solvent qualities). Based on the conformation of the DPD chains we identify the DPD equivalent transitions for linear polymers in different solvents. The maximum level of coarse graining $\phi_{max}$ that can be applied for each solvent condition is identified according to the equation \eqref{eq:sigmax}. The non-solvent limit presented is consistent with the magnitude of the spring constant $K_s$ we chose. This limit can be further exploited using stronger bond interactions or even different bond potentials.\cite{Posel2014}}
\label{fig:nuvari}
\end{figure}

The study of the conformation of single chains in solution allow us to identify how fine-grained systems are governed by the enthalpic interaction. We corroborate that DPD chains exhibit  well defined transitions from good to poor solvent,  and the theta condition occurs at $a_{ps} \approx 27.5$. Based on these results we can consistently characterize the variation of $\nu'$ with the interaction parameter. Moreover, from equation \eqref{eq:sigmax} we compute how the  maximum level of coarse graining $\phi_{max}$ varies with the polymer-solvent affinity at the fine scale (Figure \ref{fig:nuvari}). The variation of the maximum level of coarse graining is an important feature of the model-reduction  methodology we are introducing. This shows that for a given DPD chain it is not possible to apply an arbitrary coarse graining level when the chain is in different solvents, thus a DPD chain containing $N'$ segments can be reduced upto $\phi=N$ in poor solvents, while in  good-solvent condition this would be impossible.
\\
\\

\subsection{Validation of the coarse graining}
\begin{figure}[h!]
\centering
\includegraphics[width=0.75\linewidth]{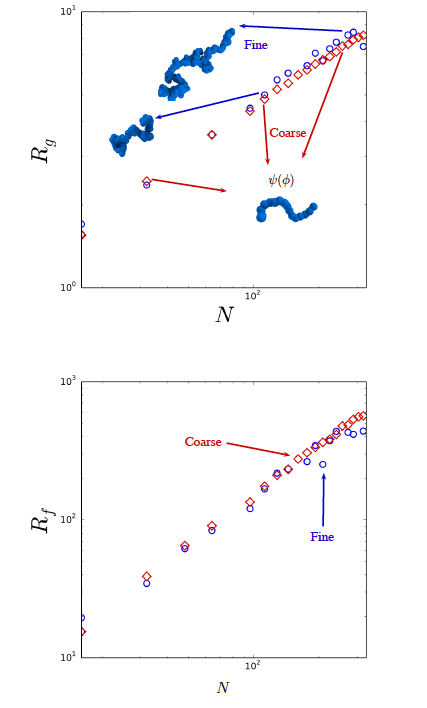}
\caption{Radius of gyration and end-to-end distance variation with molecular weight. Empty circles correspond to fine-grained chains, while rotated squares indicates the size of coarse-grained chains containing the same number of particles but different molecular weight.}
\label{fig:rgAndRfvariation}
\end{figure}

\begin{figure}[h!]
\begin{center}
\includegraphics[width=0.75\linewidth]{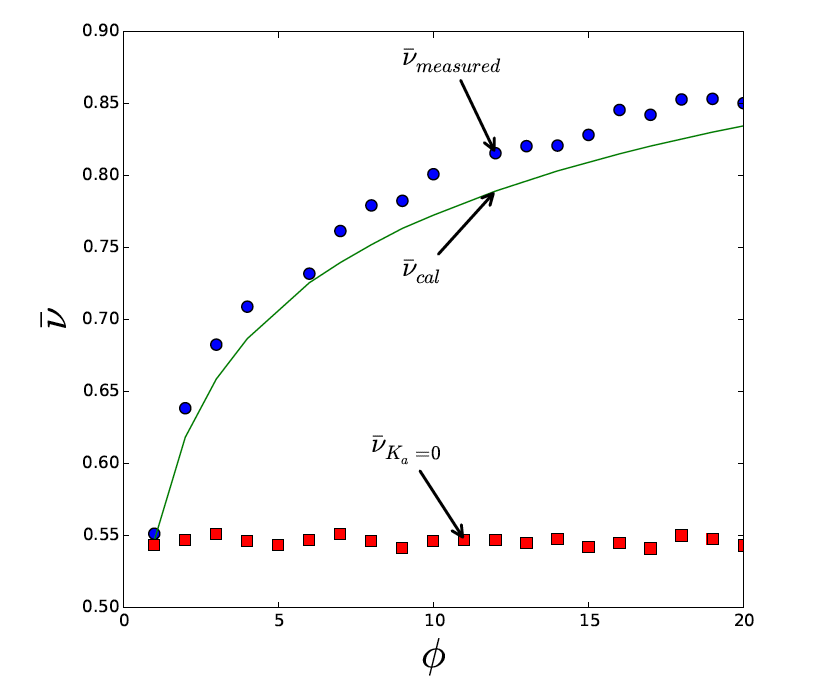}
\end{center}
\caption{Comparison between the calculated $\bar{\nu}_{cal}$ (equation \eqref{eq:nu1}) and the measured $\bar{\nu}_{measured}$ coarse-chain conformation for different levels of coarse graining. For comparison we include the variation of  the chain conformation $\bar{\nu}_{K_a = 0}$ when the entropic constraints are not imposed.}
\label{fig:theoMeasNu}
\end{figure}

Once we identify how the conformation of DPD polymer chains is driven in fine scales, and how it can be controlled in coarse models, we validate the size-preserving and forecasting capabilities of our model reduction methodology.
\\
\\
The systems evaluated contain individual DPD chains in athermal solvent ($a'_{ii}=a'_{ij}=25.0$).  We applied different levels of coarse graining $\phi$ such that all the coarse-scale chains have the same number of beads, while each represents different molecular weights. To determine the required coarse conformation $\bar{\nu}$ (equation \eqref{eq:nu1}), rather than measuring the conformation $\nu'$ for every fine-grained counterpart, we evaluate if the size of fine models can be forecast using a constant $\nu'$ derived from the study presented in Figure \ref{fig:nuvari}. In Table \ref{tab:validationDetails} we give a break-down of the parameters we use in this validation stage.
\begin{table}
 \caption {Coarse Graining over different molecular weight polymers}
  \centering

    \begin{tabular}{lr}
    \hline
    Parameter & Values                                                                    \\ \hline
    $N'$      & 16 - 320                                                         \\ \hline
    $\phi$      & 1 - 20                                                    \\ \hline
    $\bar{N}$     & 16  \\ \hline
    $a'_{ps}$     & 25  \\ \hline 
    \end{tabular}
 
 \label{tab:validationDetails}
\end{table}
\\
\\
Figure \ref{fig:rgAndRfvariation}  presents the fine-chain size variation  ($R_g$ and $R$) with the molecular weight for polymers in athermal conditions. Along with the fine-chain sizes we have included the measured radius of gyration and end-to-end distance of their equivalent coarse representations. According to the fine-scale variation identified (Figure \ref{fig:nuvari}), coarser chains are constructed taking $\nu'=3/5$. In Figure \ref{fig:rgAndRfvariation} is remarkable the agreement between fine- and coarse-grained chains in both $R_g$ and $R$. The small difference between fine and coarse models at high molecular weight appears due to deviations of the real fine conformation with respect to the value of $\nu'$ used to make the coarse models. Nevertheless, taking into account that a coarse-grained curve is constructed in a forecasting stage  taking a constant $\nu'=3/5$ the current results are satisfactory.
\\
\\
The results compiled in Figure \ref{fig:rgAndRfvariation} show that using the model reduction framework we introduce, it is possible to cover a wide range of molecular weights while using the same number of beads per chain. Thus, using geometrical considerations, our coarse-graining methodology, allows us to map fine-scale models to a reference state through a consistent system scaling $\psi(\phi)$, where short-length and fast-time scales are neglected and the relevant properties that govern the phase equilibria are preserved.    
\\
\\
To verify if the functional form \eqref{eq:alphaFunction} obtained from the reference system is independent of the level of coarse graining, and can be used to control $\bar{\nu}$, we present in Figure \ref{fig:theoMeasNu} the measured chain conformation, $\bar{\nu}_{measured}$, along with the calculated $\bar{\nu}$ (equation \eqref{eq:nu1}) for the different levels of coarse graining.  From Figure \ref{fig:theoMeasNu} we confirm that the dependence of the chain conformation with the equilibrium angle of the reference chains is consistently extended to coarse models, such that the sought $\bar{\nu}$ is properly obtained. 
\\
\\
In order to highlight the importance of the entropic constraints to preserve the relevant properties of the DPD chains, in Figure \ref{fig:theoMeasNu} we include the measured coarse-grained conformation when angle restrictions are not imposed in the model. Here it is appreciated that in absence of bending potentials, $\bar{\nu}$ is practically independent of the coarse graining level ($\nu' \approx \bar{\nu}$). From equation \eqref{eq:sigmaandnu}  ($\phi^{3\bar{\nu} - 1} = N'^{3(\bar{\nu} - \nu')}$), if the fine and coarse conformation are approximately equal there is only one conformation that satisfies the chain size and density preservation, $\nu' = \bar{\nu} = 1/3$, or $\phi=1$.  
\\
\\

\begin{figure}[h!] 
\begin{center}
\includegraphics[width=0.75\linewidth]{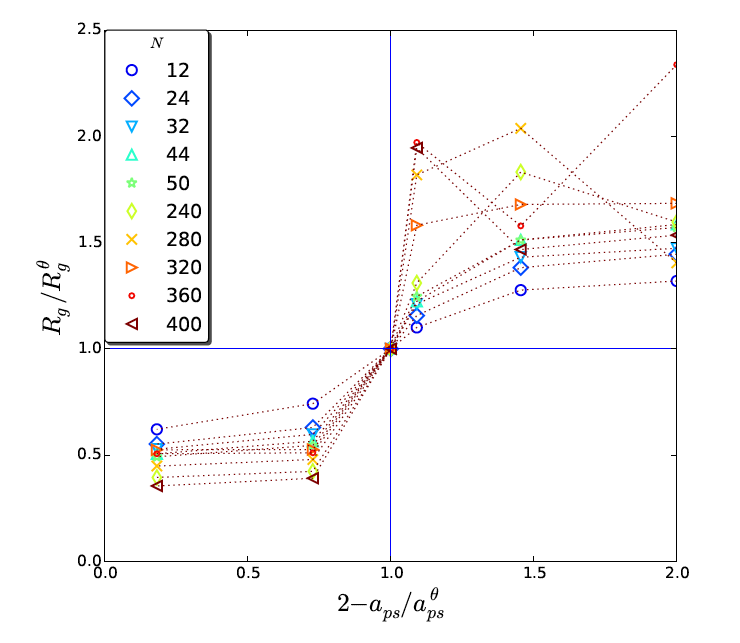}
\end{center}
\caption{Radius of gyration $R_g$ of the chain at different interaction parameters $a_{ps}$. For larger chains there is a pronounced jump in $R_g$ corresponding to the transition between theta to poor solvent.}
\label{fig:aijeffectOnRg}
\end{figure}

\begin{figure}[h!] 
\begin{center}
\includegraphics[width=0.75\linewidth]{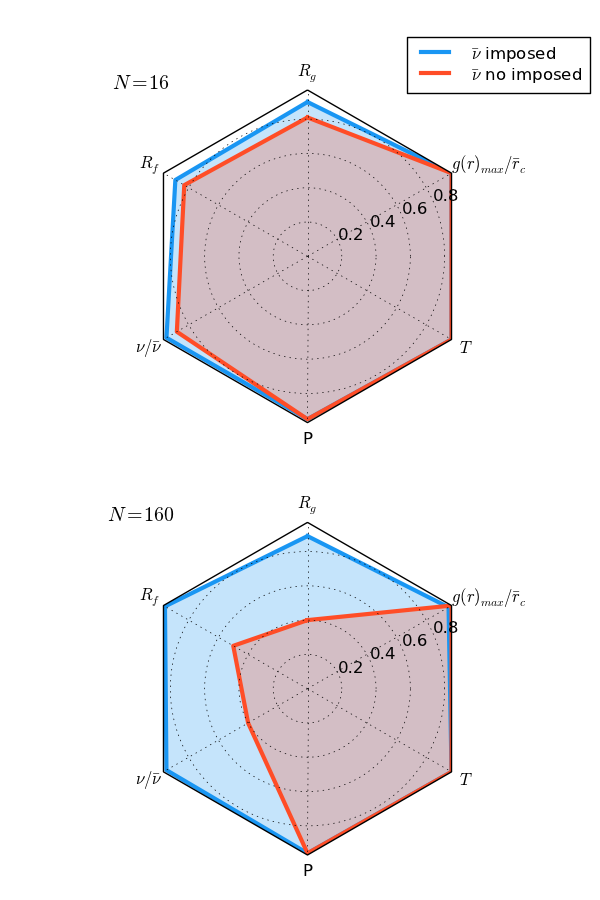}
\end{center}
\caption{Comparison of the preservation of properties after the coarse-graining procedure. For short polymer chains the impact of the entropic restrictions is not noticeable. However as the fine DPD chain grows (more coarsening is applied) only the model reduction approach proposed herein properly preserves the polymer sizes. For 16-bead chains we apply a level of coarse graining $\phi=4$, while for 160-bead chains, $\phi = 10$}
\label{fig:CGcomparison}
\end{figure}

\noindent
Despite of the inherent limitations of chain models without angle imposition, we can explain why the methodology followed by \cite{Spaeth2011} (where $\bar{\nu}$ is not adjusted) is capable to preserve fine-scale properties if the number of beads per chain  and the level of coarse graining do not exceed a maximum. On the one hand, it can be seen in Figure \ref{fig:theoMeasNu} that the difference between the conformation $\bar{\nu}$ of the restricted and unrestricted models is smaller for lower molecular weight polymers with low coarse-graining levels. Therefore when $\phi$ is small, the difference in conformation between fine and coarse-grained chains is not significant ($\nu' \approx \bar{\nu}$), and the deviations without angular restricted models are hidden. 
\\
\\
On the other hand, the limit in the number of beads per chain is observed after analysing the radius of gyration variation.  In Figure \ref{fig:aijeffectOnRg} the radius of gyration dependence, $R_g/R_g^{\theta}$, with the interaction parameter is depicted for chains of different molecular weights. Considering the variations of $\nu'$ with the solvent affinity, a noticeable jump in $R_g/R_g^{\theta}$ above the theta condition is expected and evidenced for long chain models. However, in the case of chains containing fewer particles the transition at the theta point is weaker and the change in the chain size is negligible. 
\\
\\
Finally, for the sake of consistency we  evaluate the efficacy of the coarse-graining methodology proposed, through the quotient $Q$ between the different preserved properties. Given a property \textit{A} in its fine and coarse grained representations, we compute

\begin{equation}
Q = 1-\frac{|\bar{A} - A'|}{A'},
\label{eq:effic}
\end{equation}
where $Q$ tends to one when \textit{A} is properly preserved after the coarse graining. Figure \ref{fig:CGcomparison} includes the results of the model reduction over systems containing short and long chains. In addition, we again compare the performance of the entropically-constrained coarse graining we proposed, with the coarse graining approach followed in,\cite{Spaeth2011} where the segment correlation between fine $\nu'$ and coarse $\bar{\nu}$ representations is not accounted for, yielding $\nu' = \bar{\nu}$. In the last case the interaction and spring parameters are scaled but non bond-angle potential is included. From Figure \ref{fig:CGcomparison} we identify that for short chain models (i.e., 16 bead) the effect of the entropic restrictions in the dimensions of the chain is not noticeable, and both model reduction methodologies nearly preserve the properties of the system. However for larger polymer models (i.e., 160 beads) the difference in particle correlation between fine and coarse scales becomes relevant, and the model reduction without explicit control over $\bar{\nu}$ fails.    
\\
\\

\section{Conclusions}

The model-reduction framework we describe satisfactorily preserves the relevant properties that define the phase-equilibria in polymer-solvent systems , such as the pressure, temperature, density and size ratio between species, regardless the length of the DPD chain and the level of coarse graining. However the explicit dependence of the coarse graining with the chain conformations imposes limits in the maximum level of coarsening that can be achieved. The methodology proposed can be widely applied to different particle-based method, in particular, we present our validation  in the context of Dissipative Particle Dynamics (DPD).


\bibliographystyle{unsrt}
\bibliography{references}

\end{document}